\documentclass{aa}  

\usepackage{graphicx}
\usepackage{amsmath}
\usepackage{amsfonts}
\usepackage{txfonts}
\newcommand{\um}{\,$\mu$m }
\newcommand{\amax}{$a_{\rm max}\,$}

\usepackage{booktabs,array}

\begin{document}

   \title{Exploring the dust grain size and polarization mechanism in the hot and massive Class 0 disk IRAS 16293-2422 B}

   \titlerunning{Grain sizes and dust polarization in IRAS 16293 B}
   \authorrunning{Joaquin Zamponi et al.}
   
   \author{Joaquin Zamponi\inst{1}, 
          Mar\'ia Jos\'e Maureira\inst{1},
          Hauyu Baobab Liu\inst{2},  
          Bo Zhao\inst{1}, 
          Dominique Segura-Cox\inst{1,3,\thanks{NSF Astronomy and Astrophysics Postdoctoral Fellow}}, 
          Chia-Lin Ko\inst{4}, 
          \and
          Paola Caselli\inst{1}
          }

   \institute{
            Max-Planck-Institut für extraterrestrische Physik, Gießenbachstraße 1, D-85748, Garching bei München, Germany; \\
              \email{jzamponi@mpe.mpg.de}
        \and
            Department of Physics, National Sun Yat-Sen University, No. 70, Lien-Hai Road, Kaohsiung City 80424, Taiwan, R.O.C.
        \and
            Department of Astronomy, The University of Texas at Austin, 2500 Speedway, Austin, TX, 78712, USA
        \and
            Department of Astronomy, University of Arizona, 933 N. Cherry Ave., Tucson, AZ, 85721, USA
             }

   \date{}

 
  \abstract
   {
   Multiwavelength dust continuum and polarization observations arising from self-scattering have been used to investigate grain sizes in young disks.   
   However, the likelihood of self-scattering being the polarization mechanism in embedded disks decreases for very highly optically thick disks and puts some of the size constraints from polarization on hold, particularly for the younger and more massive disks.
   The 1.3\,mm polarized emission detected toward the hot ($\gtrsim$400 K) Class 0 disk IRAS 16293-2422 B has been attributed to self-scattering, predicting bare grain sizes between 200-2000\,$\mu$m.
   }
   {
   We aim to investigate the effects of changing the maximum grain sizes in the resultant continuum and continuum polarization fractions from self-scattering for a hot and massive Class 0 disk extracted from numerical simulations of prestellar core collapse and compare with IRAS 16293 B observations.
   }
   {
   We compared new and archival dust continuum and polarization observations at high-resolution between 1.3 and 18\,mm to a set of synthetic models. 
   We have developed a new public tool to automate this process called Synthesizer. This is an easy-to-use program to generate synthetic observations from numerical simulations.
   }
   { 
   Optical depths are in the range of 130 to 2 from 1.3 to 18\,mm, respectively. 
   Predictions from significant grain growth populations, including $a_{\rm max}=1000\,\mu$m are comparable to the observations from IRAS 16293 B at all observed wavelengths. 
   The polarization fraction produced by self-scattering reaches a maximum of $\sim$0.1\% at 1.3\,mm for a maximum grain size of 100\,$\mu$m, which is an order of magnitude lower than that observed toward IRAS 16293 B.
   }
   {
   From the comparison of the Stokes I fluxes, we conclude that significant grain growth could be present in the young Class 0 disk IRAS 16293 B, particularly in the inner hot region ($<10\,$au, $T>$ 300 K) where refractory organics evaporate.
   The polarization produced by self-scattering in our model is not high enough to explain the observations at 1.3 and 7\,mm, and effects like dichroic extinction or polarization reversal of elongated aligned grains remain other possible but untested scenarios. 
   }


   \maketitle
%

\section{Introduction}
\label{sec:Introduction}
Circumstellar disks, the sites of planet formation, are relatively well-studied in the more evolved Class II stage. 
However, in the earliest Class 0 phase, protostellar disks have only recently begun to be studied in detail (e.g., \citealt{Maury2019,Segura-Cox2018,Tychoniec2020,Zamponi2021}), and many of their properties such as mass, temperature and dust properties remain unknown, or highly debated. 
Since Class 0 protostellar disks set the initial conditions for planet formation, which could be underway already in the Class I phase \citep{Sheehan2018, Segura-Cox2020}, constraining the sizes of dust grains in disks at the earliest time possible is critical. 
This is usually done by constraining the variations of the spectral index at millimeter wavelengths, which can be related to the dust opacity index \citep{Testi2014}. 
In addition, polarization observations have proven to be a useful independent method to constrain the grain sizes in disks. This can be done if the origin of the polarized emission is self-scattering \citep{LazarianAndHoang2007a, LazarianAndHoang2007b, Kataoka2015, Andersson2015, Tazaki2017}.
Self-scattering refers to the polarization of the dust thermal emission scattered by the dust itself, and produces emission polarized to a few percent, at millimeter wavelengths for grain sizes in the micro- and millimeter range \citep{Kataoka2017}. 
The amount of polarization depends on the level of anisotropy of the radiation field generated by the self-scattered photons of the dust emission, which in turns depends on the inclination of the disk and the size of the scattering dust particles \citep{Yang2016a}.   
The relation between the size of the dust grains and the maximum polarization fraction turns this polarized emission into a tool to estimate the level of grain growth at different evolution stages.   

In the more evolved (Class II) protoplanetary disks, optical and near-infrared observations of scattered flux usually trace the emission from small dust grains that scatter off the protostellar flux in the upper layers of the disk \citep{Avenhaus2018, Garufi2018, Garufi2019, Garufi2020}. At millimeter wavelengths, the observed dust scattered emission is mainly originated from the dust thermal emission \citep{Kataoka2015,Yang2016a}.

In this work we explore the effect of changing the maximum grain sizes in the resultant continuum and polarization fractions by dust self-scattering at millimeter wavelenghts. We use a model produced by a radiation-hydrodynamic (RHD) numerical simulation of core collapse. The disk formed is massive ($\sim$0.3 M$_{\odot}$), hot ($>$300 K, within 10 au), and optically thick at millimeter wavelenths. This disk model successfully reproduced Stokes I fluxes from ALMA $\sim$6-10 au resolution observations, at 1.3 mm and 3 mm, for the very young Class 0 disk IRAS 16293 B \citep{Zamponi2021}. Thus we also aim to test whether we can reproduce the fluxes and levels of polarization fraction observed in this source when the polarization is produced by self-scattering of spherical grains in the Mie regime. 
We expand on our previous exploration on grain sizes in this young disk \citep{Zamponi2021} by showing the effects of varying the maximum grain size (\amax) in the model from 10\,$\mu$m to 1000\,$\mu$m. Furthermore, as the disk is hot enough to evaporate solid organics, we show the effects of evaporating water and potentially some organics within the so-called sootline and having grain growth within this region, motivated by recent laboratory \citep{Gundlach2018, Pillich2021} and observational \citep{Liu2021b} results. 

IRAS 16293-2422 B is a well studied Class 0 protostar located in the star-forming region $\rho$-Ophiuchi, and it is one of the closest and brightest protostars at a distance of 141 pc \citep{Dzib2018}. 
The disk around the protostar is hot ($T_{\rm b}\gtrsim400\,$K) and massive, and likely subject to  gravitational instabilities, as initially proposed by \citet{Rodriguez2005} and \citet{Dipierro2014}, and recently confirmed by \citet{Zamponi2021}.  
The source is very young ($<$10$^{4}$\,yr, \citealt{Andre1993}) and its water snow line extends over a 20\,au radius \citep{Zamponi2021}, making it an ideal laboratory to test the scenario proposed by recent laboratory experiments \citep{Gundlach2018,Pillich2021,LiAndBergin2021}, that grain growth is boosted in dry conditions.  
The disk mass estimated by \citet{Zamponi2021} is 0.003\,$M_{\odot}$ in solid material, which would be 33 times higher than the minimum mass solar nebula of 30$M_{\oplus}$ (in solids; \citealt{Weidenschilling1977_MMSN, Andrews2020ARA&A}).  
This implies that IRAS 16293B contains enough mass to form super-Earth planetesimals. 
Because of all these conditions, IRAS 16293B represents one of the ideal sites to probe grain growth at the earliest stages. 

In the younger Class 0 disks, although polarization by self-scattering at millimeter wavelengths have also been proposed (e.g., \citealt{Sadavoy2018, Tsukamoto2022PPVII}), studies simulating the feasibility of this mechanism in more realistic young disk models, such as those formed out of the collapse of a core using 3D numerical simulations have not been done. This is particularly important since the physical properties of deeply embedded disks appear different in numerical simulations (e.g., \citealt{Zamponi2021, Xu2021a,Xu2021b,Bate2022}) as compared to widely used analytical models for Class II disks \citep{Ballering2019}. 

In IRAS 16293-2422 B, polarized light was initially observed by  \citet[][]{Rao2009,Rao2014} with the SMA at 0.87\,$\mu$m, who detected polarization fractions of around $\sim$1.4\%, distributed mostly azimuthally around the protostar location. 
The resolution of this detection was 0.6" (85\,au) and likely traced envelope scales, which correspond to the bridge structure connecting the northern and southern protostars \citep{Jorgensen2016,Maureira2020}.  
Additional observations were presented by \citet{Liu2018b} with the VLA at 7\,mm, and a 1.5 times better resolution. 
These observations resolved down to 50\,au and found a similar polarization pattern and fraction ($\lesssim$2\%).
More recently, the survey of 1.3\,mm polarization observations toward Class 0 protostars in $\rho$-Ophiucus, carried out by \citet{Sadavoy2019} at a 2 times better resolution (0.2"; 30\,au), found polarization signatures associated to self-scattering in inner regions of most of their sources. 
The polarization in the disk of IRAS 16293 B was found to be azimuthal, similar to the SMA and VLA observations, however the authors associated the vector distribution to polarized self-scattering from an optically thick face-on disk, based on the models from \citet{Yang2017}. 
The connection of a similar polarization pattern between 1.3\,mm and 7\,mm, being produced by self-scattering, implies that grains can have sizes between 200-2000\,$\mu$m. 
In this work, we use a disk model from numerical simulations to test this hypothesis.

This paper is structured as follows: in section \ref{sec:Observations} we provide details on the archival and new observational data used in this work, in section \ref{sec:dust_model} we describe the dust model and polarization scheme used for the radiative transfer analysis, in section \ref{sec:results} we present the results of the assessment of grains sizes through the stokes I fluxes and our self-scattering models, in section \ref{sec:Discussion} we discuss grain growth and possible mechanisms responsible for the observed polarization and other scenarios, and finally, in section  \ref{sec:Conclusions} we present the conclusions of this work.

\section{Observations}
\label{sec:Observations}

\subsection{ALMA \& VLA archival data}
\label{sec:ALMA_VLA_archival_data}
We have compiled archival multiwavelength polarization observations of IRAS 16293-2422 B at 1.3\,mm (ALMA Band 6), 3\,mm (ALMA Band 3) and 7\,mm (JVLA Band Q). 
The 1.3\,mm data used in this work is twofold. 
For the analysis presented in section \ref{sec:constraint_grain_sizes} we used high-resolution Stokes I continuum images, with a resolution of 0.114"$\times$0.069" and a noise level of 104$\,\mu$Jy\,beam$^{-1}$ (0.3 K; see colorscale from leftmost panel in Fig. \ref{fig:observations}).  
For the analysis of polarized data, we used publicly available Stokes I, Q and U images by \citet{Sadavoy2018}, each with a resolution of 0.18"$\times$0.09" and a noise level of 280$\,\mu$Jy\,beam$^{-1}$ (0.4 K), 25$\,\mu$Jy\,beam$^{-1}$ and 25$\,\mu$Jy\,beam$^{-1}$, respectively.  
Band 6 polarization vectors shown in Fig. \ref{fig:observations} have been masked in regions were the stokes I flux is lower than 3$\sigma_{\rm I}$ and the polarized intensity is lower than 3$\sigma_{\rm Q}$. 
The 3\,mm data we used is only available in stokes I, but presented the highest resolution image of IRAS 16293 B currently available, with a beam size of 0.048"$\times$0.046" and a noise level of 17$\,\mu$Jy\,beam$^{-1}$ (0.95 K). 
The data used at 7\,mm contains full polarization information, with a resolution of 0.39"$\times$0.24" and a noise level of 35$\,\mu$Jy\,beam$^{-1}$ (0.25 K). 
Polarization vectors have been masked following the same criteria as for Band 6 observations. 

We refer the reader to \citet{Zamponi2021} for more information about the high-resolution ALMA observations at bands 6 and 3 as well as their calibration and imaging. 
Similarly, we refer to \citet{Liu2018b} for the details on the observations, calibration and imaging of the polarized data at the JVLA band Q.

\subsection{VLA Band Ka \& Ku observations}
\label{sec:VLA_BandKu_obs}
We present JVLA standard continuum mode observations at Ka (9\,mm) and Ku band (18\,mm) toward IRAS 16293-2422 B, both in the A array configuration. 
Band Ka observations were carried out on March 7, 2022 (project code: 22A-322, PI: Zamponi). 
Band Ku observations were carried out on January 2, 04, and 05 2021 (project code: 20B-172, PI: Chia-Lin Ko). 
The pointing and phase referencing centers for our target source are R.A. $=$ 16$^{\mbox{\scriptsize{h}}}$32$^{\mbox{\scriptsize{m}}}$22$^{\mbox{\scriptsize{s}}}$.610 (J2000), decl. $=$ $-$24$^{\circ}$28$'$32$\farcs$61 (J2000) for our Band Ka observations, and R.A. $=$ 16$^{\mbox{\scriptsize{h}}}$32$^{\mbox{\scriptsize{m}}}$22$^{\mbox{\scriptsize{s}}}$.620 (J2000), decl. $=$ $-$24$^{\circ}$28$'$32$\farcs$5 (J2000) for Band Ku. 
We employed the 3 bit sampler in both observations.

For the band Ka observations, 24 antennas were used with a projected baseline range of 794-34.4 kilometers. 
Bandpass, phase and amplitude calibrators used were J1256-0547, J1625-2527 \& J1331+3030, respectively.
We calibrated the data using the standard VLA Pipeline (v2021.2.0.128) using the Common Astronomy Software Applications (CASA; \citealt{McMullin2007}) package, release 6.2.1.7. 
The spectral configuration included 10 narrow (16 MHz) and 52 wide (128 MHz) spectral windows. 
We created our continuum image using wide windows only, with a total bandwidth of 7 GHz, centered at 33 GHz. 
We performed phase-only self-calibration iteratively in CASA. 
In each iteration the cleaning was done with the deconvolver \texttt{mtmfs}, a robust parameter of 0.5 and scales 0, 1 and 3 beams. 
The final image has a resolution of 0$''$.12$\times$0$''$.05 (P.A.$=$-23.11$^{\circ}$) and a noise level of 16$\mu$Jy\,beam$^{-1}$ (3 K).

For the band Ku observations, the absolute flux/passband and complex gain calibrators were 3C286 and J1625-2527.
The projected baseline range is $\sim$1.15-36.50 kilometers.
We manually followed the standard data calibration strategy using CASA (release 5.6.2). 
We utilized the built-in image model for 3C286 during the calibrations.
After implementing the antenna position corrections, weather information, gain-elevation curve, and opacity model, we bootstrapped delay fitting and passband calibrations, and then performed complex gain calibration. 
We applied the absolute flux reference to our complex gain solutions, and then applied all derived solution tables to the target source. 
Finally, we based our observations on 3C286 to solve the cross-hand delay and absolute polarization position angles, and took J2355+4950 as a low polarization percentage calibrator when solving the leakage term (i.e., the D-term).
We performed the zeroth order (i.e., nterm$=$1) multifrequency synthesis imaging. 
The Briggs Robust$=$0 weighted image achieved a noise level of 47 $\mu$Jy\,beam$^{-1}$ and a 0$''$.23$\times$0$''$.094 (P.A.$=$5.2$^{\circ}$) synthesized beam.

\begin{figure*}
    \centering
    \includegraphics[width=18cm]{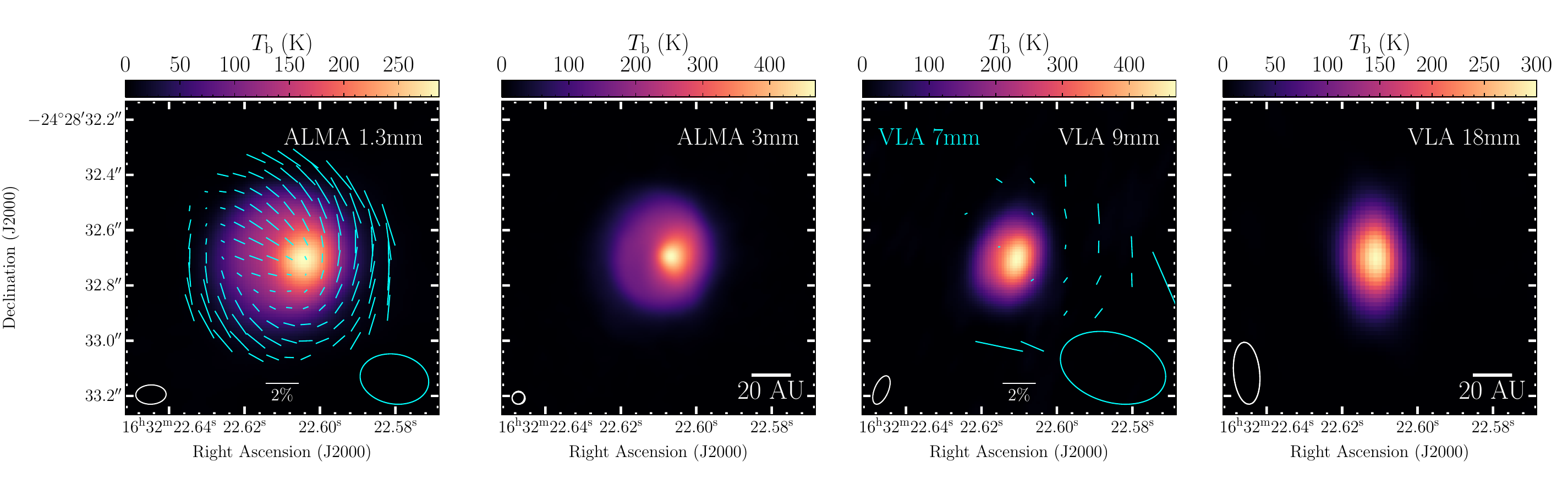}
    \caption{IRAS 16293-2422 B observed at ALMA bands 6 \& 3 and VLA bands Q, Ka \& Ku, from left to right. Cyan vectors represent polarization E-vectors (masked out below 3$\sigma_{\rm I}$ and 3$\sigma_{\rm Q}$) and the scalebar indicates a level of 2\% polarization fraction. The continuum data at 1.3 and 3\,mm was presented by \citet{Zamponi2021} and the 1.3\,mm polarization observations were presented by \citet{Sadavoy2018}. The polarized continuum observations at 7\,mm were obtained from \citet{Liu2018b}. 9\,mm and 18\,mm continuum observations are presented in this work for the first time. The angular resolution of the observations are: 0.114"$\times$0.069" at 1.3\,mm (0.18"$\times$0.09" for polarized data), 0.048"$\times$0.046" at 3 mm, 0.39"$\times$0.24" at 7\,mm, 0.12"$\times$0.05" at 9\,mm and 0.23"$\times$0.09" at 18\,mm.}
    \label{fig:observations}
\end{figure*}

Finally, since all the observations shown in Fig. \ref{fig:observations} have been taken at different times, we have been corrected the images for the proper motion of the source calculated based on previous observations with the VLA \citep{Hernandez-Gomez2019b} and the 3 mm observations presented here. 
The applied correction is RA: $-11.8 \pm 0.3$ mas/yr and DEC: $-19.7 \pm 1.3$ mas/yr. 
We have aligned the observations to the observing time of the most recent band Ka observation (2022/03/07).

\section{Disk and dust model}
\label{sec:dust_model}
\subsection{Protostellar disk model}
\label{sec:disk_model}

\begin{figure}
    \centering
    \includegraphics[trim=1cm 1cm 0cm 1cm, clip, width=\columnwidth]{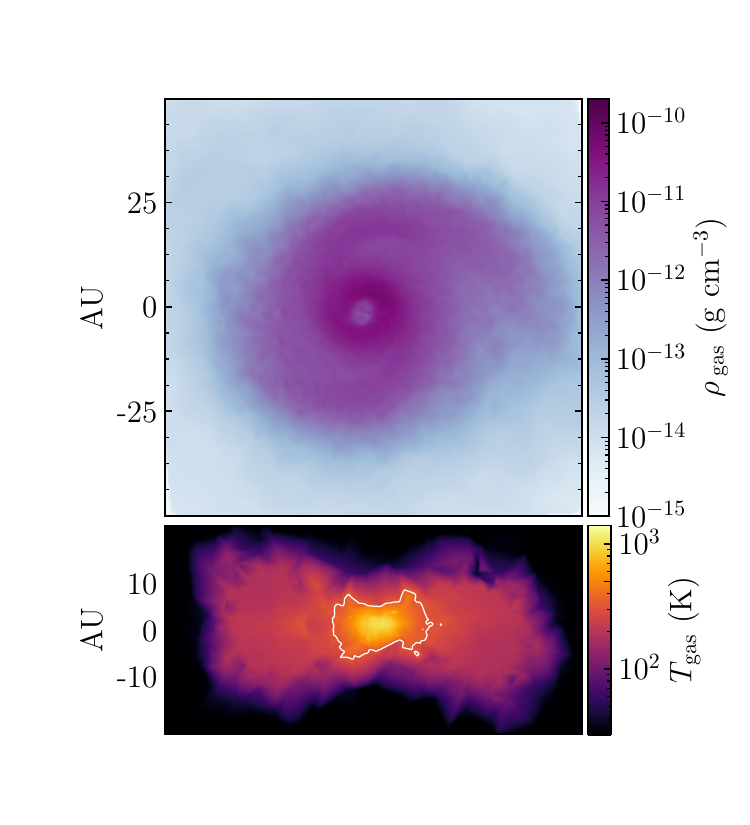}
    \caption{Face-on gas density (top) and edge-on temperature (bottom) distributions across the midplane of the protostellar disk  formed from the numerical simulations collapse of a core ans that we used to compare to the observations of IRAS 16293-2422 B. The white contour in gas temperature represents the extension of the soot-line at 300 K. This model is presented in greater detail in \citet{Zamponi2021}.}
    \label{fig:disk_model}
\end{figure}

For comparison with the observations of the IRAS 16293-2422 B disk, we use a numerical simulation of the collapse of a prestellar core which results in a hot and gravitationally unstable protostellar disk. This simulation has been presented in \citet{Zamponi2021} and the resultant density and temperature structure is also shown in Fig. \ref{fig:disk_model}.  
In \citet{Zamponi2021} we showed that this disk can successfully reproduce both the 1.3 and 3 mm observed fluxes and the spectral index between both bands. 
This implies producing high brightness temperatures (e.g., $T_{\rm b}^{\rm 3mm}\gtrsim400\,$K at the peak) to match those from IRAS 16293B observed with ALMA.  
The disk is obtained from a simulation snapshot at a stage 18.2 kyr after the collapse of a 1\,M$_{\odot}$ spherical and isothermal cloud, simulated using the smoothed-particle-hydrodynamics (SPH) code \texttt{sphNG} \citep{Bate1995}. 
This simulation setup included a radiation transport scheme. 
This helps to more accurately reproduce the cooling and heating source terms along its temporal evolution and to deliver more realistic gas temperature distributions.   
The mass in the disk at this point is $\sim$0.3\,M$_{\odot}$.  
The gas temperature is $\gtrsim$300\,K within the central 10\,au and decreases to $\sim$200\,K at the scales where two roughly symmetric spiral arms have formed (10-30\,au; see Fig. \ref{fig:disk_model}). 

To generate the radiative transfer model, we have set the dust temperature to be equal to the gas temperature from the RHD simulation. 
This is justified because of the high densities found within the disk, that allow the dust to be dynamically and thermally coupled with the gas (see also \citealt{Zamponi2021}). 
We assume a homogeneous gas to dust density ratio of 100. 
We have used the RADMC3D radiative transfer code \citep{radmc3d} with a regular cartesian grid, generated by interpolating all particle positions. The particle interpolation and gridding was done with our newly developed tool, called Synthesizer (see section \ref{sec:synthesizer}), which automates the generation and execution of radiative transfer models from numerical simulations. 
The resulting dynamic range and spatial distribution of the gas density and temperature between our gridding scheme and that from \citet{Zamponi2021} are in very good agreement. 
The former gridding scheme used in \citet{Zamponi2021} consisted on a Voronoi tessellation of the particle locations which was supported in POLARIS \citep{Reissl16}, but it is not yet fully supported in RADMC3D (beta feature).

\subsection{The Synthesizer: from simulations to synthetic observations}
\label{sec:synthesizer}
In order to automate the comparison of observational data and simulation outputs, we have developed a new tool called \textsc{Synthesizer}, which we have made publicly available\footnote{\url{https://github.com/jzamponi/synthesizer}} in the form of a python package\footnote{\url{https://pypi.org/project/astro-synthesizer}}. 
Synthesizer is a program to calculate synthetic images from either an analytical model or numerical simulations directly from the command-line. 
For SPH simulations, it interpolates the particle positions into a rectangular cartesian grid and then uses RADMC3D to do the Monte-Carlo and raytracing. 
Then it feeds the output image to CASA to generate a final synthetic observation. 
Support for polarization models, either by scattering or grain alignment, is also included.  
Additionally, the \textsc{Synthesizer} includes a module called the \textsc{DustMixer}.   
This is a tool to generate dust opacity tables and full scattering matrices from the optical constants of a given material, all from the command-line. \textsc{DustMixer} also allows to experiment with the mixing of different materials and different grain sizes. 
For further information about this dust module, see appendix \ref{app:dustmixer}.

\subsection{Grain sizes}
\label{sec:dust_grain_sizes}
In this work we have used a dust model similar to that used by \citet{Zamponi2021}, corresponding to spherical grains with a minimum size of 0.1\,$\mu$m. 
This lower end of the size distribution is chosen based on the prediction for removal of very small grains during the stage of protostellar disk formation \citep{Zhao2018, Silsbee2020}. 
In this work we explore maximum grain sizes in IRAS 16293B by showing the effect of $a_{\rm max}$ on the resultant simulated fluxes.  
In \citet{Zamponi2021} we modeled the emission of IRAS16293B using a dust population with $a_{\rm max}=10\mu$m only, which successfully reproduced the observed brightness temperatures. 
The millimeter scattering opacities in that case are negligible (see Fig. \ref{fig:dust_opacities}). 
The results in \citet{Zamponi2021} demonstrated that the low ($<\sim$2) 1.3-3mm spectral index observed in the IRAS 16293 B disk could be reproduced without the need for larger grains in which the scattering opacities are important. 
The reason for this was that the resultant disk from the numerical simulation is both hot toward the inner layers and optically thick at millimeter wavelengths, both necessary conditions to recover the low observed alpha (see Figure 9 in \citealt{Zamponi2021}). 
Here we extend the analysis by considering distributions with $a_{\rm max}=10, 100$ \& $1000\,\mu$m, and compare the resulting images to the observations presented in Fig. \ref{fig:observations} (see section \ref{sec:constraint_grain_sizes}). 

The dust opacities computed for different models are presented in Fig. \ref{fig:dust_opacities} (silicates + graphites and silicates + graphites + organics).  
The figure shows three panels, one for each maximum grain size, and tabulated in Table \ref{tab:opacities} for the 4 observing wavelengths from Fig. \ref{fig:observations}.  
For each $a_{\rm max}$, three different linestyles represent absorption, scattering and extinction opacities for a given composition (colored lines).  
When comparing the three panels, we can see at a glance the relation between the albedo ($\kappa_{\rm sca} / \kappa_{\rm ext}$) and the maximum grain size.
The wavelength regime in which $\lambda\sim2\pi a_{\rm max}$ is also where the albedo is the highest.   
Because of this, the millimeter scattering opacities can vary by orders of magnitude for maximum grain sizes between 10 and 1000\,$\mu$m.  
These differences reflect also on the level of polarized scattered flux (produced by dust self-scattering) and turn it into a proxy of grain growth \citep{Kataoka2016a, Kataoka2016b, Yang2016a, Stephens2017, Harris2018, Sadavoy2018, Ohashi2020, Lin2021}.

\subsection{Composition}
\label{sec:dust_composition}
\begin{figure*}
    \centering
    \includegraphics[width=20cm, trim=2.1cm 0 0.0cm 0, clip]{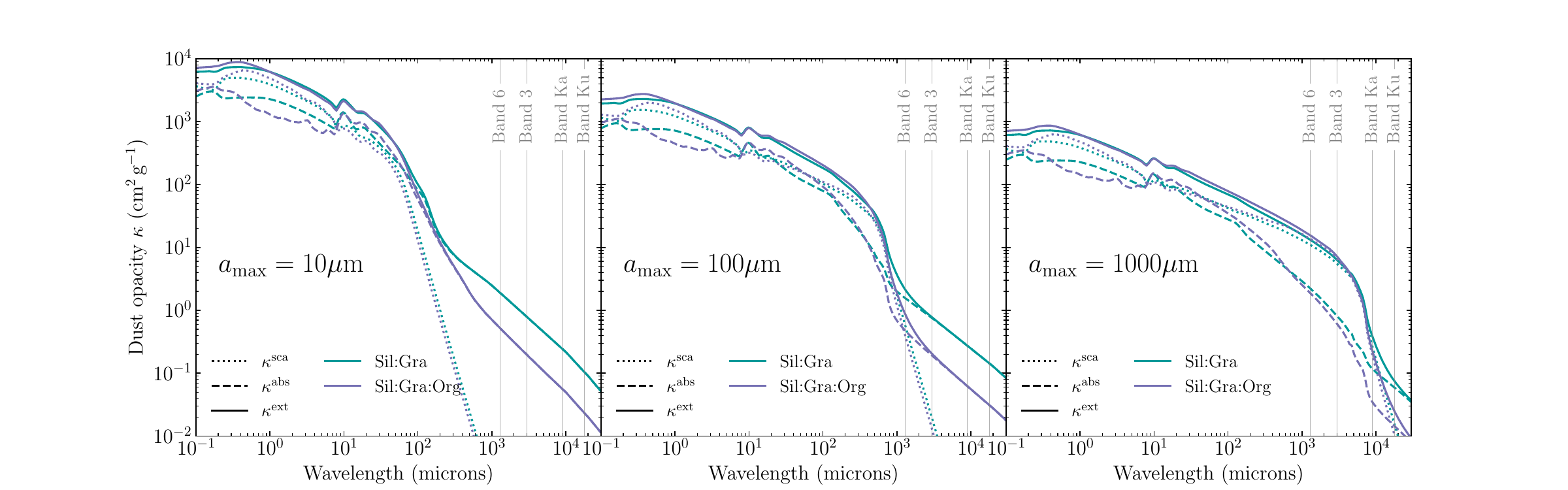}
    \caption{Dust opacities for different maximum grain sizes and compositions. The three panels show, from left to right, dust opacities for a maximum grain size of 10, 100 \& 1000\,$\mu$m, respectively. Each panel contains opacities generated for a mixture of silicates and graphite (0.625-0.375) and a mixture of silicates, graphites and refractory organics (0.627-0.300-0.075; 80\% of organics). The gray curve on every panel indicates the albedo of the fiducial composition (silicates and graphites). Four vertical gray lines indicate the observing wavelengths used in this study, corresponding to the observations from Fig. \ref{fig:observations}. The exact opacity values used in this work are listed in Table \ref{tab:opacities}, for different $a_{\rm max}$ and compositions, at all four wavelengths.}
    \label{fig:dust_opacities}
\end{figure*}

Our fiducial dust mixture consists of 62.5\% of astronomical silicates and 37.5\% of graphite (Sil:Gra curve in Fig. \ref{fig:dust_opacities}).
The optical constants\footnote{\url{https://github.com/jzamponi/synthesizer/tree/main/synthesizer/dustmixer/nk}} for both materials were obtained from \citet{Draine2003a} and \citet{Draine2003b}, respectively.  
However, carbon-bearing species in the interstellar medium may be present in many forms and shapes, and may not necessarily be all crystallized as graphite. 
They can also be amorphous or ring-like (as in PAHs) carbonaceous.
The predominant shape of carbonaceous material in space is hard to determine \citep{Jager1998, Zubko2004, Birnstiel2018DSHARP}.    
According to \citet{Draine2003a}, molecules with sizes larger than 0.1\,$\mu$m are likely to be crystallized. 
This is in fact the lower end of our size distribution (see section \ref{sec:dust_grain_sizes}), hence, we assume graphite as a fiducial representation of the carbon budget in our dust model.   

To consider the effects of carbon evaporation, we have also considered inclusions of refractory organics \citep{Henning1996} into the carbon mixture and studied the resulting differences in the overall opacity.
Refractory organics have considerably lower sublimation temperature ($\sim$300\,K, \citealt{Jager1998, vantHoff2020a, LiAndBergin2021}) compared to that of graphite ($\sim$2100\,K). 
Taking into account the lower sublimation temperatures of certain carbonaceous materials is important because they decrease the dust mass in hot regions and results in an overall decrease in the dust opacity. 
This sublimation zone within which the amount of carbon in the grains is reduced, also called "soot-line" \citep{Kress2010, vantHoff2020a}, is likely to be present in IRAS 16293B at a radius of $\sim$10\,au, based on the high ($T_{\rm b}\gtrsim400$\,K) brightness temperatures observed (see Fig. \ref{fig:observations}). 
The spatial extension of this region in the simulated disk is illustrated by the white contour in the edge-on temperature projection in Fig. \ref{fig:disk_model} which extends radially up to $\sim$10\,au and vertically to $\sim$8\,au. 

The difference between opacities with and without organics can be seen in Fig. \ref{fig:dust_opacities}.  
In the two compositions with carbon, the total carbon fraction of 37.5\% is kept constant.  
For the one including refractory organics, we replaced 50\% of the graphite by refractory organics, meaning, both have mass fractions of 18.75\%.  
At millimeter wavelengths, the opacity of a dust population with \amax of 10 or 100\um\, is largely dominated by graphite instead of silicate. 
When we compare the two different compositions, i.e., silicates-graphites versus silicates-graphites-organics, we see that a mixture with inclusions of organics has a lower opacity than a purely graphitic material. 
This is because the opacity of pure graphite is larger than that of pure organics and including organics into the mixture implies a partial removal of graphite, since our carbon budget is maintained constant\footnote{We acknowledge that the presence of
refractory organics in the icy mantles of dust grains should in principle not affect the content of core carbon (i.e., graphite) and that a proper sublimation model should evaporate the refractory carbon without affecting the graphitic core. However, the nature of the mixing process in our setup (see section \ref{sec:synthesizer}), forces us to reduce the mass fractions of some materials to account for the inclusion of new others, while keeping the dust mass constant.}. 
When considering organics evaporation in our radiative transfer calculation we change the dust compositions accordingly. 
In regions where $T_{\rm dust}<300\,$K, the dust is a mixture of silicates, graphites and refractory organics, whereas in regions where $T_{\rm dust}\geq300\,$K ($\sim$10\,au radius), we removed the organics from the mixture to mimic the sublimation of carbon and locally scaled down the dust-to-gas mass ratio by the corresponding mass-loss factor. 
Within the soot-line, the organics get sublimated and removed from the dust grains.
This produces a reduction of the dust mass and opacity which results in a reduction of the optical depth.  

Two major features can be concluded from Fig. \ref{fig:dust_opacities}: (I) for every $a_{\rm max}$ the difference in opacity between both compositions is very small, and (II) in the regime where $\lambda\sim 2\pi a_{\rm max}$, i.e., where the albedo is maximum (Mie regime), the dust opacity is highly insensitive to variations in the dust composition. 
This means that observations of polarized emission produced by scattering can serve as a tool to constrain grain sizes but not compositions. 
The small differences we find for different compositions are associated with the resulting extinction opacity and not with the polarization pattern or fraction. 
We acknowledge that in some cases, variations in the scattering polarization patterns can indeed be produced by different dust compositions, as it is shown by \citet{Yang2020}.   
Their works show that compositions like those from \citet{Kataoka2015} or \citet{Yang2016a} can produce polarization reversal (i.e., 90 degrees rotation of the vector angles) when elongated millimetric grains are present.

\begin{table}[]
    \centering
    \begin{tabular}{l|l|c|c|c|c}
                              & \! & \!Band 6 & \!Band 3 & \!Band Ka & \!Band Ku\\ 
          \,\,\,$a_{\rm max}$ &    \! Mix       & \!1.3 mm & \!3 mm   & \!9 mm    & \!18 mm\\ 
                          & & & & & \\ 
        \hline
                          & & & & & \\ 
           \!\!10 $\mu$m  & s-g &  1.84 & 0.76 & 0.24 & 0.10 \\ 
                          & s-g-o &  0.52 & 0.20 & 0.06 & 0.02\\ 
                          & & & & & \\ 
          \!\!100 $\mu$m  & s-g &  2.10 & 0.76 & 0.27 & 0.14 \\ 
                          & s-g-o &  0.86 & 0.20 & 0.06 & 0.03 \\ 
                          & & & & & \\ 
          \!\!1000 $\mu$m & s-g   & 13.19 & 6.22 & 0.38 & 0.07 \\ 
                          & s-g-o & 15.68 & 7.13 & 0.22 & 0.02 \\ 
                          & & & & & \\ 
    \end{tabular}
    \caption{Dust opacities (g cm$^{-2}$) used in this work, for different maximum grain sizes, different compositions and at four wavelengths. The two compositions are: silicate with graphite (s-g) with mass fractions 0.625 and 0.375, respectively, and silicate with graphite and refractory organics (s-g-o), with mass fractions 0.625, 0.1875 and 0.1875 (80\% of carbon), respectively. }
    \label{tab:opacities}
\end{table}

\subsection{Polarization by dust self-scattering}
\label{sec:polarized_scattering}
We have performed radiative transfer calculations in all four Stokes components (I, Q, U, V) using the radiative transfer code \textsc{RADMC3D} \citep{radmc3d} via the Synthesizer.     
To model the Stokes I fluxes produced, we raytraced the dust thermal emission along the line-of-sight and included the emission scattered by the dust grains, both from stellar and dust thermal radiation. 
The dominant source of scattering at millimeter wavelengths is the scattered dust thermal emission, namely, self-scattering \citep{Kataoka2015}. 

Scattering of light in dusty media is commonly modeled as a stochastic process of absorption and re-emission of light  \citep{Bjorkman&Wood2001} by means of a Monte-Carlo (MC) simulation \citep{Steinacker2013}. 
The scattering event is modeled as the re-emission of an absorbed photon in a new random direction. 
The likelihood for a given direction is not isotropically distributed, but rather follows the commonly used phase function from \citet{HenyeyAndGreenstein1941}. 
However, this kind of scattering model considers only information about the light intensity and not of the polarization state, i.e, stokes I only flux, without stokes Q and U.  
To calculate polarized flux, the scattering process has to be represented by a matrix rotation of all four Stokes components. 
This process is also called full Mie scattering \citep{Mie1908,BohrenAndHuffman,WolfAndVoschinnikov2004} and is significantly more computationally expensive than the case for the Stokes I only.
To ensure both approaches (i.e., Stokes I only and Full Mie) deliver similar Stokes I fluxes, we tested the convergence of the scattering Monte-Carlo. The convergence was reached at about 10$^{10}$ photons. 

The information associated with linear polarization of light is stored in the Q and U components of the Stokes vector.  
We estimate the total linearly polarized intensity $PI$, polarization degree $P_{\rm frac}$ and polarization angle $P_{\rm angle}$ as 
\begin{equation}
    \label{eq:polarized_intensity}
    PI = \sqrt{Q^2 + U^2},
\end{equation}
\begin{equation}
    \label{eq:polarization_fraction}
    P_{\rm frac} = \frac{PI}{I}, 
\end{equation}
and 
\begin{equation}
    \label{eq:polarization_angle}
    P_{\rm angle} = \frac{1}{2}\arctan\left(\frac{U}{Q}\right), 
\end{equation}
respectively.

\subsection{Synthetic observations}
\label{sec:synthetic_observation}
We post-processed the output from the radiative transfer with a series of synthetic observations, using the observing setups of the ALMA and VLA observations shown in Fig. \ref{fig:observations}. 
This step was performed using the Synthesizer (see section \ref{sec:synthesizer}). 
Synthesizer uses the CASA (v5.6.2) software and its \textsc{simobserve} and \textsc{tclean} tasks to produce a synthetic image from a model image, created by the output of RADMC3D.  
The CASA scripts for the setup at every band are available within the Synthesizer public repository\footnote{\url{https://github.com/jzamponi/synthesizer/tree/main/synthesizer/synobs/templates}}.   

All synthetic maps include the thermal noise associated to the corresponding observing setup (time, bandwidth, cycle, etc.), and lead to noise levels comparable to the real observations.   
In regions of the images where Q and U are not well detected, equation (\ref{eq:polarized_intensity}) leads to a positive bias in the polarized intensity that must be removed \citep{SimmonsAndStewart1984, Vaillancourt2006}. 
The polarized intensity must be then debiased by taking into consideration the thermal noise in the polarization map $\sigma_{\rm PI}$ as

\begin{equation}
    \label{eq:polarization_debias}
    PI = \sqrt{Q^2 + U^2 - \sigma^2_{\rm PI}},
\end{equation}
where we assume that $\sigma_{\rm Q} \sim \sigma_{\rm U} \sim \sigma_{\rm PI}$ \citep{Vaillancourt2006}. 
When producing maps of polarized emission and polarization vectors from synthetic observations, we mask out the vectors at locations where $\mathrm{I}/\sigma_{\rm I}<3$ and $PI/\sigma_{\rm PI}<3$, same as we did for the real observations in Fig. \ref{fig:observations}.

\section{Results}
\label{sec:results}
\subsection{Effects of different maximum grain sizes on the Stokes I fluxes}
\label{sec:constraint_grain_sizes}
\begin{figure*}
    \centering
    \includegraphics[width=18cm, trim=1.8cm 1cm 1.5cm 1cm, clip]{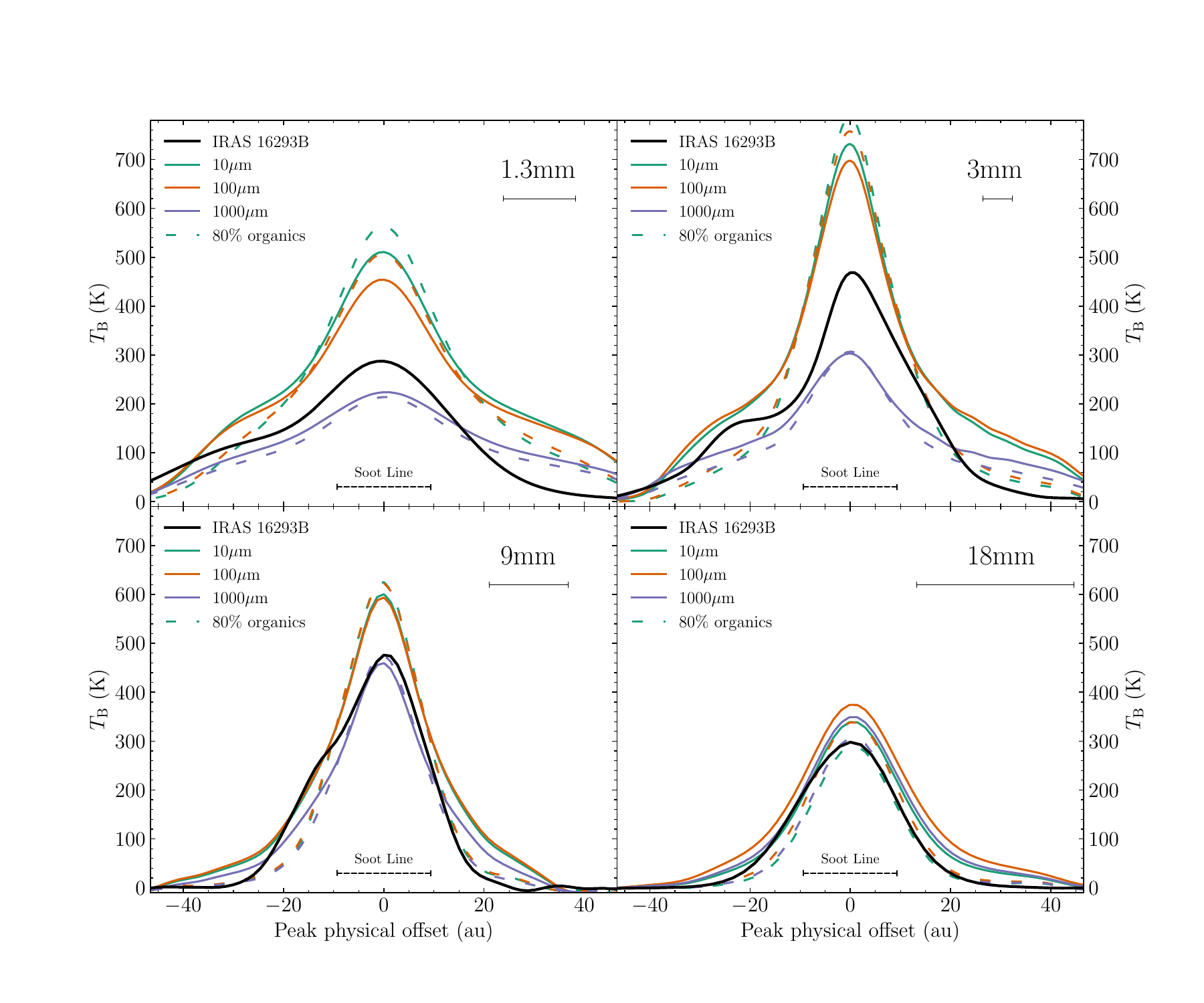}
    \caption{Cuts of the brightness temperature distribution along the east-west axis for real and synthetic observations at four wavelengths. The black solid line in each panel represents the ALMA and VLA observations shown in Fig.~\ref{fig:observations}. Colored lines indicate models with different maximum dust grain size, taking values of $a_{\rm max}$= 10, 100 \& 1000\,$\mu$m. We have included models accounting for the sublimation of 80\% of the carbonaceous material, in the form of refractory organics. The sublimation zone extends over a 10\,au radius where the gas temperature exceeds 300\,K. The black line under every wavelength label, indicates the angular resolution.}
    \label{fig:horizontal_cuts}
\end{figure*}

We start our analysis by producing synthetic observations of the Stokes I fluxes from the simulated disk at 4 different wavelengths: 1.3, 3, 9 \& 18\,mm, corresponding to the ALMA bands 6 \& 3 and VLA bands Ka \& Ku (Fig. \ref{fig:observations}). 
All Stokes I images and profiles presented in this paper represent net fluxes. This means thermal plus scattered flux.  
For each band we mimicked the observing setup of the real observations described in section \ref{sec:Observations} and presented in Fig. \ref{fig:observations}. 
We extracted a cut of the brightness temperature maps along the east-west axis and plotted it as a function of physical offset from the peak, as shown in Fig. \ref{fig:horizontal_cuts}.  
All brightness temperature cuts are taken along the position of the peak flux which correspond to offset zero in Fig. \ref{fig:horizontal_cuts}.  
The black line in Fig. \ref{fig:horizontal_cuts} represents the observed source brightness, and the three different colored solid lines represent the models with different $a_{\rm max}$. 
A black horizontal line in the upper right corner of each panel indicates the geometric mean between the major and minor axis of the beam.   
Dashed lines show models with carbon-sublimation (see section \ref{sec:carbon_sublimation}).

\begin{figure*}
    \centering
    \includegraphics[width=18cm, trim=1.8cm 1cm 1.5cm 1cm, clip]{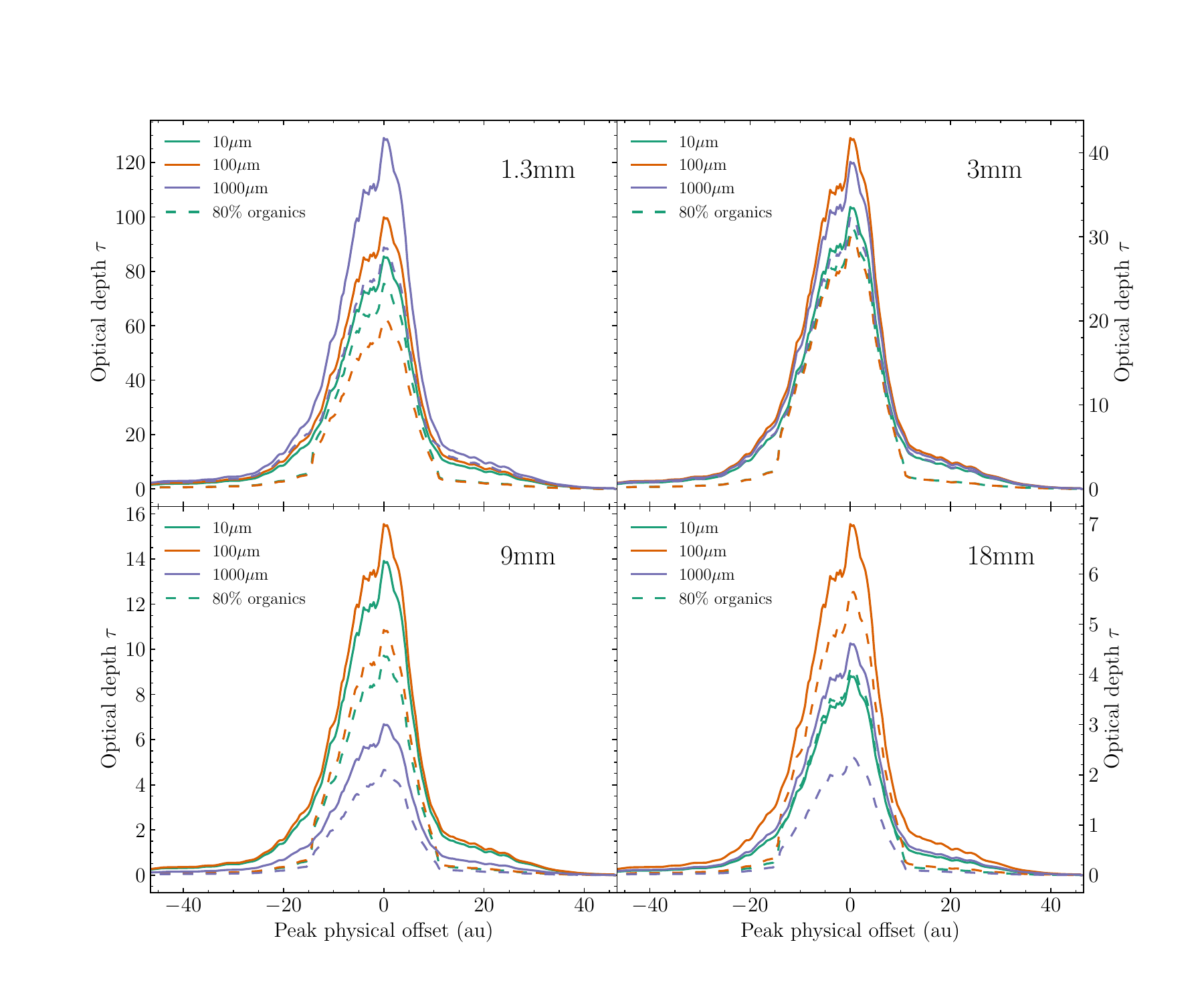}
    \caption{Similar to Fig. \ref{fig:horizontal_cuts} but for the optical depth of every radiative transfer model at the ideal resolution, this is, at the resolution of the radiative transfer output before any beam convolution.}
    \label{fig:horizontal_cuts_optical_depth}
\end{figure*}

At each wavelength, the flux decreases with increasing $a_{\rm max}$. 
This is because all the models are optically thick (see Fig. \ref{fig:horizontal_cuts_optical_depth}), and the larger the \amax the higher the optical depth. 
This effect, along with the decreasing temperature as a function of the scale height, leads to colder dust temperatures being traced by the larger $a_{\rm max}$. 
For \amax= 100\um and 1000$\mu$m, the fluxes are further lowered from the dust temperature of the $\tau=1$ layer due to a significant albedo \citep{Birnstiel2018DSHARP}. 
This also explain why for a given $a_{\rm max}$, the peak fluxes increase with wavelength as longer wavelength penetrate further within the disk where the temperatures are higher. 
We note that at 18 mm both the observed models and observations are affected by beam dilution. 
The net fluxes from both \amax= 10 and 100 $\mu$m are similar because they are completely determined by the extinction optical depth, which is similar between these two models (see Fig. \ref{fig:horizontal_cuts_optical_depth}). 
The difference between these two models lies on the albedo, which is higher for $a_{\rm max}=100\,\mu$m.

When comparing our models to the observations (black profile in Fig. \ref{fig:horizontal_cuts}), a similar trend is seen. 
This is a feature of a optically thick disks with a positive temperature gradient toward the center.
In this central regions, the observed Class 0 disk fluxes lie between the models with $a_{\rm max}=100\,\mu$m and 1000\,$\mu$m, with differences between the observations and models within a factor of up to 2.  
On the outskirts, most models tend to overpredict fluxes up to a factor of 2 toward negative offsets and a factor of several for positive offsets. 
This east-west difference is because our disk model does not have such a marked east-west asymmetry as the one observed in IRAS 16293 B (see Fig. \ref{fig:observations}). 
Scenarios that can explain the observed asymmetry like the presence of asymmetric spiral arms or an off-center protostar were discussed in \citet{Zamponi2021}.  
The comparison between models and observations suggest that grains could have grown significantly in this young Class 0 disk even up to millimeter grain sizes, provided that the vertical gradients of temperature and density are close to those in IRAS 16293 B.

\subsection{Effect of carbon sublimation within the soot line}
\label{sec:carbon_sublimation}
Because the observed brightness temperatures and model dust temperatures are high ($\gtrsim400$ K), it might be that some of the dust material gets evaporated at these early stages. 
Silicate sublimation temperature is around 1200 K, and for graphite it is 2100 K. 
Both well above the central disk temperatures. 
However, carbon can also exist in nongraphitic forms, like amorphous carbon or Polyciclyc Aromatic Hydrocarbons (PAHs) within ice layers. 
To study the scenario of dust containing refractory carbon that can sublimate at the observed disk temperatures, we also produced synthetic observations with a dust model that includes amorphous carbon. 
For this we use the optical constants from amorphous carbon (CHON) from \citet{Henning1996} and set a sublimation temperature of 300\,K. 
This temperature is based on the analysis presented by \citet{LiAndBergin2021} who estimate a temperature range between 200 and 650\,K from meteoritic constraints. 
This range was further narrowed to 300\,K by \citet{vantHoff2020a} after analyzing the relation between sublimation, gas temperature and pressure at hot-core conditions. 
The thermodynamic relations they used were derived from PAHs characterized in the lab \citep{GoldfarbAndSuuberg2008,Siddiqi2008}.      
In our disk model, the soot-line at 300\,K covers the inner 10\,au in radius and $\sim8$ au in scale height (see Fig. \ref{fig:disk_model}).  
The extension of this region is also shown with a black scalebar at the bottom of Fig. \ref{fig:horizontal_cuts}. 
Details on the dust composition and radiative transfer setup for the inclusion of amorphous carbon are given in section \ref{sec:dust_composition}. 

The resulting brightness temperature profiles for the case with carbon sublimation are also shown in Fig. \ref{fig:horizontal_cuts} in the form of dashed lines. 
In the sublimation case the grain size remains homogeneous but the composition does not. 
Outside of the sootline, grains are composed by silicate, graphite and refractory organics (opacity given by the purple line of Fig. \ref{fig:dust_opacities}). 
Inside the sootline, amorphous carbon is removed and the composition is the fiducial silicate and graphite mixture (turquoise line in Fig. \ref{fig:dust_opacities}), however, with a dust mass reduced by the factor of material sublimated. In Fig. \ref{fig:horizontal_cuts} the dashed lines show the the case of 80\% organics, meaning a dust composition whose carbon budget was split between 20\% of graphite and 80\% of amorphous carbon, while keeping the total carbon budget constant at 37.5\%, i.e., with mass fractions of 0.675, 0.300 and 0.075 for silicate, graphite and amorphous carbon, respectively. We tried different percentages for organics but we show here only the one with the most significant differences from our fiducial composition. 

The fluxes produced by sublimated grains are not significantly different to those without sublimation, regardless of the size and wavelength or particular amount of organics.   
As discussed in section \ref{sec:dust_composition}, the sublimation has the effect of lowering the optical depth because of the mass reduction and the change in composition.  
This results in observed emission being produced by a slightly inner and hotter parcel of the disk, which leads to slightly higher fluxes toward the center than the fiducial model without organics evaporation. 
The difference in flux in the central region between models with and without evaporation is larger for the wavelenghts with higher resolution observations.

Our results indicate that detecting dust sublimation with continuum observations at a particular wavelength, even with very high-resolution, is challenging. 
Future observations that can obtain molecular line emission from this region, can probe whether an active carbon-rich chemistry exists or not in this source \citep{vantHoff2020a}, providing independent robust constraints on the scenario of a sootline of about 10\,au within this Class 0 disk.

\subsection{Variation of the spectral index}
\label{sec:contraints_from_spectral_index_map}

\begin{figure*}
    \centering
    \includegraphics[width=18cm, trim=1.5cm 0cm 0 0, clip]{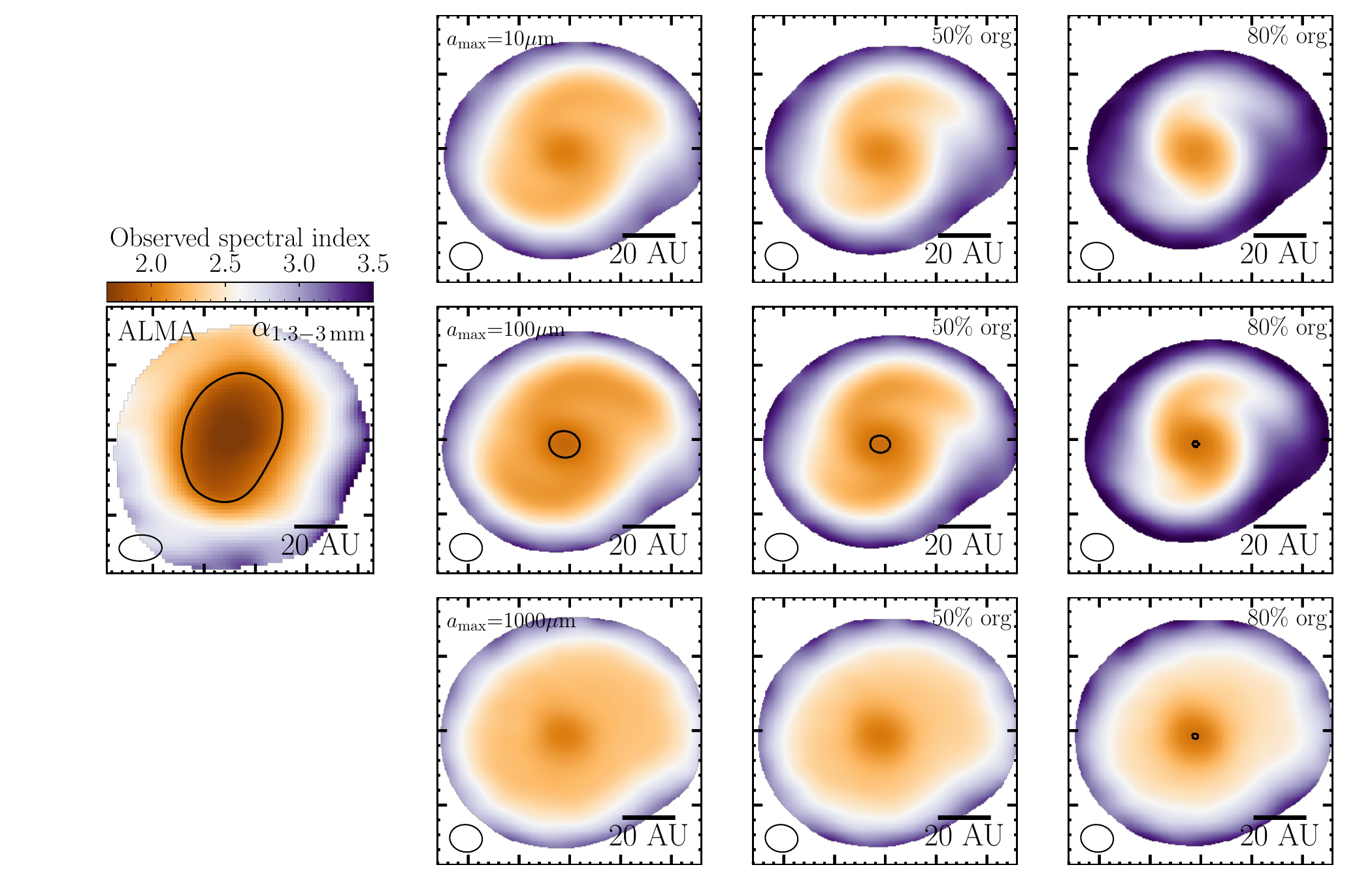}
    \caption{Spectral indexes for the real observation (left) and models (right). Models are shown for $a_{\rm max}$ of 10 (top) and 100\um (center) and 1000\um (bottom). Columns represent the cases without sublimation and with 50\% and 80\% of carbon sublimation, from left to right, respectively. The black contour indicates $\alpha=2$.}
    \label{fig:spectral_indexes}
\end{figure*}

Another observable piece of comparison between observations and models is the spectral index $\alpha$ produced between 1.3 and 3\,mm, for which there are higher resolution and deeper observations available.   
We computed the spectral index between 1.3 and 3\,mm for all $a_{\rm max}$ models and for several percentages of organics, 10\%, 30\%, 50\% and 80\%. 
We present in Fig. \ref{fig:spectral_indexes}, our results for models with our fiducial composition, where carbon does not sublimate, and representative sublimation models with 50\% and 80\% of organics with sublimation temperatures from 300\,K.  

Since the disk model is optically thick at 1.3 mm and 3 mm (see Fig. \ref{fig:horizontal_cuts_optical_depth}), the values of the spectral index will depend on the vertical distribution of temperature and opacity in the disk. 
The models in Fig. \ref{fig:spectral_indexes} have all the same temperature gradient but the opacity changes with \amax and composition leading to the observed differences. 
The models with higher opacities toward the center show a more extended region with $\alpha<2.5$ values.

An important feature observed in the IRAS 16293 B spectral index is that central values go as low as 1.7, this means, lower than the threshold of $\alpha=2$ (indicated by the black contour in Fig. \ref{fig:spectral_indexes}) for optically thick emission, first presented in \citet{Zamponi2021}. 
The presence and extent of this feature in our models depend on having sufficient variations of temperature along the line of sight between the layers traced by the different wavelenghts (with $T_{\rm 1.3 mm} < T_{\rm 3 mm}$). 
The larger the region for which these differences are present the more prominent and widespread these feature is. 
In the case of 100\,$\mu$m the feature is more prominent because of the higher optical depths while in the case of 1000\,$\mu$m sized grains with carbon sublimation, the feature is likely present because the reduced optical depth in the inner hot region allows flux to come from a region with a larger gradient in temperature than the same models with less amount of sublimation. 
Although none of our models reproduce the extent of this region, models with grain growth are able to reproduce small regions showing this feature (panels with 100\,$\mu$m or 1000\,$\mu$m considering carbon sublimation).

\subsection{Polarization by self-scattering}
\label{sec:self-scat_models}

\begin{figure*}
    \centering
    \includegraphics[width=17cm]{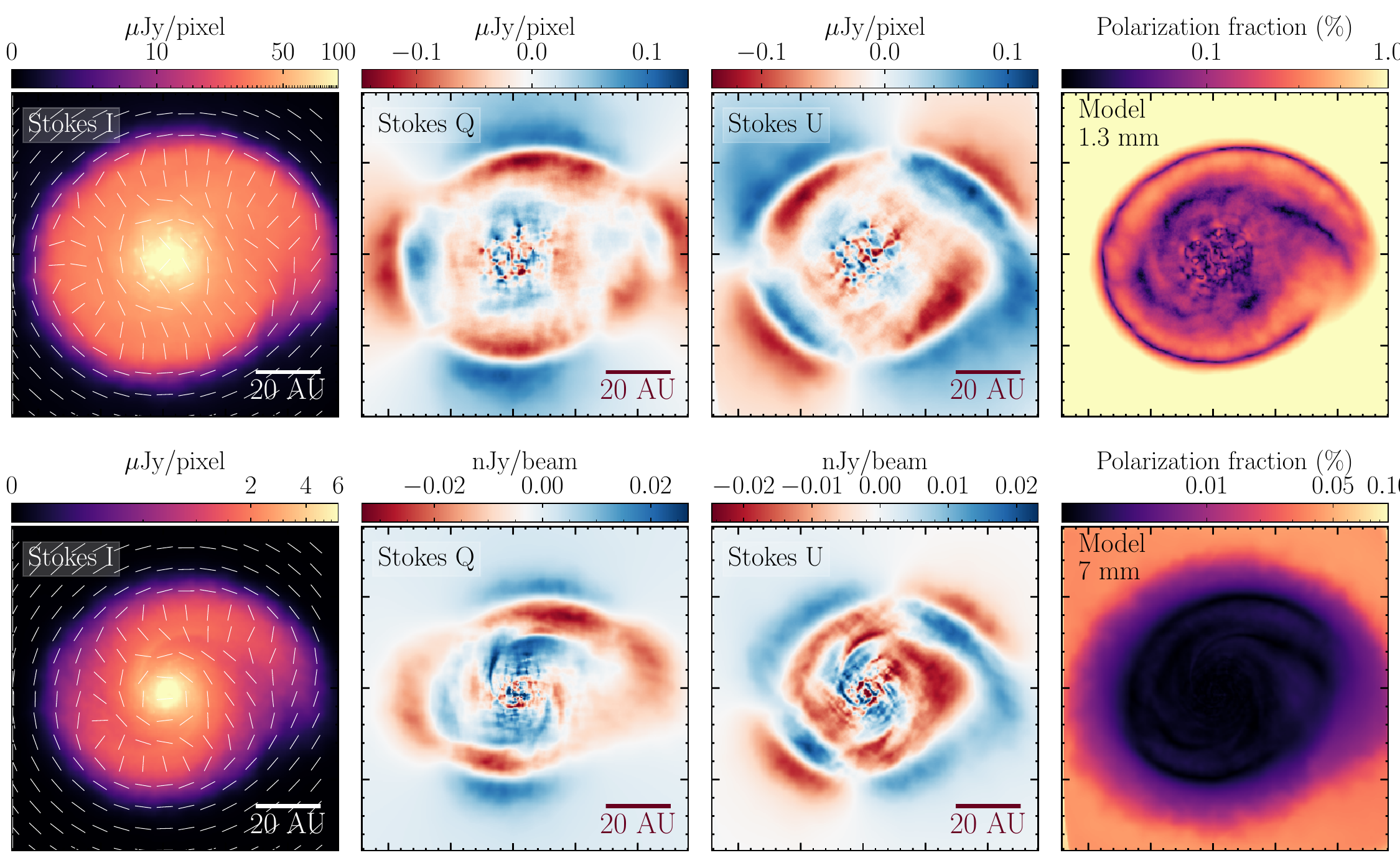}
    \caption{Radiative transfer models of self-scattering at 1.3 (top) and 7\,mm (bottom) for $a_{\rm max} = 100\,\mu$m, shown in the upper and lower row, respectively. Panels illustrate the Stokes I, Q, U fluxes and polarization fraction (with constant-size polarization vectors overlaid) from left to right.}
    \label{fig:self-scat_models_radmc3d}
\end{figure*}

Polarization observations toward embedded disks have been interpreted as caused by self-scattering, similar to that observed toward more evolved Class I sources \citep{Kataoka2016a, Stephens2017, Lin2021}. 
This interpretation is based on the morphology of the polarization observations.   
In this work, we test whether the disk model we use in this work, with a physical density and temperature structure representative of a Class 0 disk, can produce the few percent (2-4\%) levels of polarization fractions as observed for IRAS 16293 B as well as other Class 0/I disk observations at millimeter wavelengths \citep{Lee2021}.  
We do so by generating radiative transfer models of polarization by scattering at 1.3 and 7\,mm, meant to be compared with the available high-resolution polarization data shown in Fig. \ref{fig:observations}. 
Creating models of polarization by scattering requires the calculation of full Mie scattering (i.e., Monte-Carlo scattering in all four stokes components) and is highly computationally expensive. 
Hence, we produced models for all three maximum grain sizes, using only our fiducial and homogeneous dust composition of silicates and graphites. 
Given that the albedo between the different compositions used in this work are very similar, the results at each $a_{\rm max}$ will not deviate significantly when considering the other compositions. 
Between the three models, the highest polarized intensity is produced by an \amax of 100\,$\mu$m, since the albedo of 10$\,\mu$m grains is negligible and the extinction opacity of 1000$\,\mu$m sized grains larger than for 100\,$\mu$m sized grains (see Table \ref{tab:opacities}) and produces lower scattered flux. 
Unlike for \amax = 10 and 100$\,\mu$m, the polarization fraction of millimetric grains is very similar at both 1.3\,mm and 7\,mm, both in polarization pattern and intensity (maxmimum $P_{\rm frac}\sim$\,0.2\%). 
This is because their albedo is almost constant at all millimetric wavelengths (see Fig. \ref{fig:dust_opacities}). 
The results of our self-scattering models are presented in Fig. \ref{fig:self-scat_models_radmc3d} at the disk model's native resolution. 
We present the resulting Stokes I, Q and U fluxes, along with the polarization fraction (and polarization E-vectors overlaid) at both wavelengths.
The resulting Stokes Q and U fluxes at 1.3\,mm are extremely low.  
The polarization fraction in the model reaches around a tenth of percent ($P_{\rm frac}\lesssim0.1\%$) within the central 20\,au  where it shows an azimuthal polarization pattern for the E-vectors.  
This azimuthal distribution is expected from a centrally concentrated density distribution \citep{Kataoka2015}, 
in the spiral arms, the vectors become radial or rather perpendicular to these structures. 
These patterns are similar to those presented by \citet{Kataoka2015} for a ringed and lopsided disk.   
The fraction of polarization falls significantly at 7\,mm as expected, since it is far from the $\lambda\sim 2\pi a_{\rm max}$ regime, and barely reaches 0.005\% within the disk. 
We also tested the results for the case of an edge-on hot disk (Fig. \ref{fig:appendix_edge-on_disk}). 
In this case the polarization fraction remains the highest at 1.3\,mm but consistently low ($\sim1$\%).

\begin{figure*}
    \centering
    \includegraphics[width=15cm]{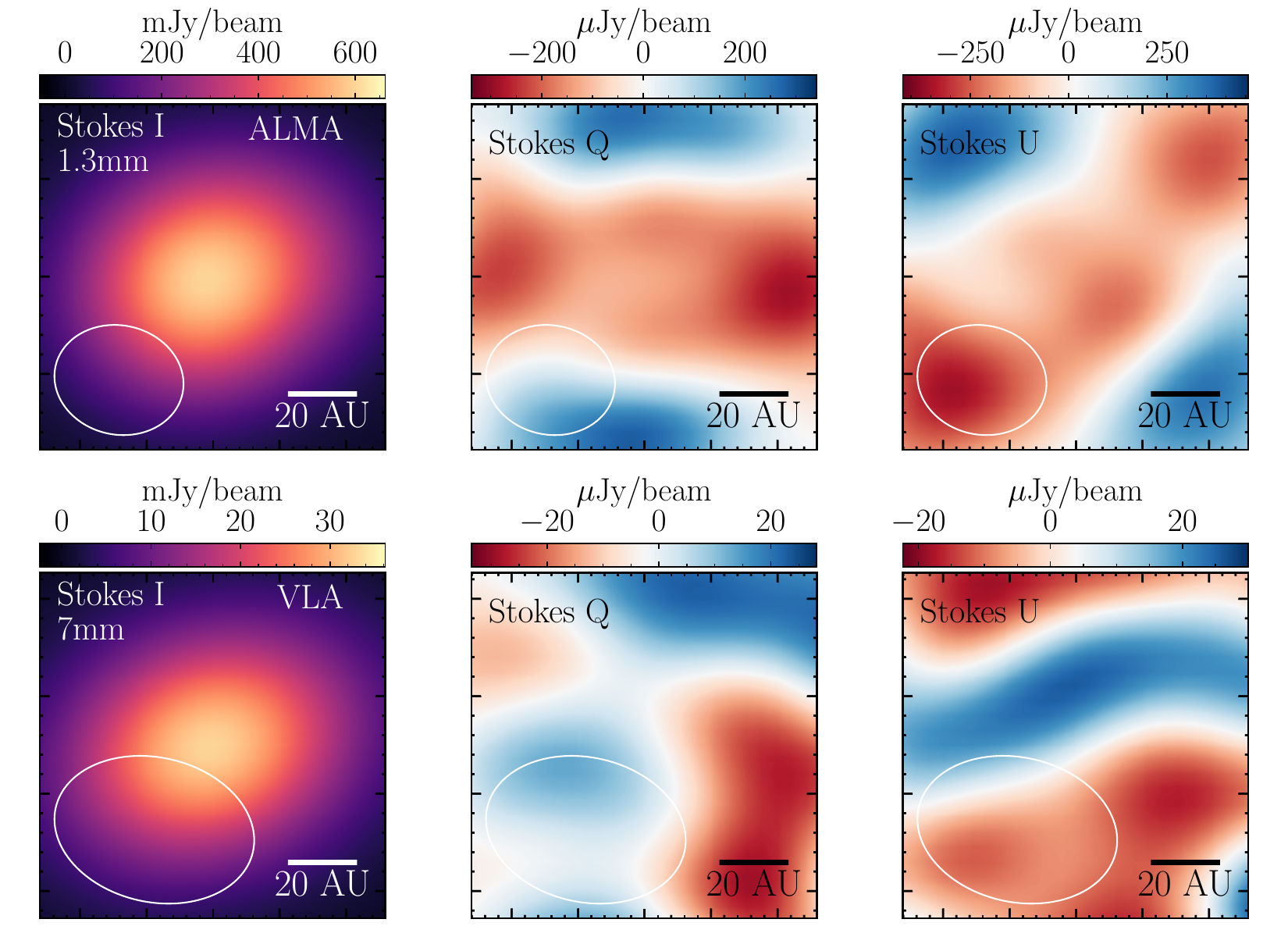}
        \caption{ALMA and VLA synthetic observations based on models of self-scattering from Fig.~\ref{fig:self-scat_models_radmc3d} (i.e., for $a_{\rm max} = 100\,\mu$m). The polarization fraction panel is omitted since both Q and U components are dominated by thermal noise. The rms of the polarization components are $rms_{\rm\, Q}\sim$ $rms_{\rm\, U}\sim$ 25\,$\mu$Jy\,beam$^{-1}$ and $\sim$12\,$\mu$Jy\,beam$^{-1}$, for the 1.3\,mm and 7\,mm maps, respectively.}
    \label{fig:self-scat_models_obs}
\end{figure*}

Previous ALMA Band 6 polarization observations of IRAS 16293 B from \citet{Sadavoy2018} show an azimuthal E-vector pattern, with polarization fractions as high as $\sim2-4$ \%. 
The authors have associated this polarization emission to that produced by dust scattering from the dust thermal emission itself (i.e., self-scattering). 
To directly compare these models with IRAS 16293 B observations, we performed polarized synthetic observations at all three Stokes components, mimicking the corresponding observing setups presented in section \ref{sec:Observations}, and the procedure described in sections \ref{sec:synthetic_observation}.  
The resulting synthetic maps (in units of Jy/beam) are presented in Fig. \ref{fig:self-scat_models_obs} for both wavelengths and all three Stokes components. 
In this case, we have omitted the polarization fraction results from the layout because the polarized intensity is completely dominated by thermal noise. 
The resulting Stokes I fluxes are comparable to the original observations, confirming the analysis presented in section \ref{sec:constraint_grain_sizes} but now with the inclusion of the polarized 1.3\,mm data.  
However, the model Stokes Q and U fluxes are not detectable within the sensitivity levels achieved by these observing setups. 
We conclude from this analysis, that linear polarization produced by self-scattering from spherical grains in our model is not high enough to be consistent with the observations of IRAS16293B carried out at 1.3 and 7\,mm by \citet{Sadavoy2018} and \citet{Liu2018b}, respectively. 

Our resulting polarization fractions are in-line with the predictions for IRAS 16293 B discussed by \citet{Yang2016a} from analytical models of polarized scattering in optically thick media. 
The millimeter optical depth in IRAS 16293 B is extremely high ($\tau\gtrsim$100), which reduces the degree of anisotropy in the radiation field and hence the percentage of polarized scattered light \citep{Kataoka2015}.

\section{Discussion}
\label{sec:Discussion}
 
\subsection{Grain growth in Class 0 disks}
\label{sec:comparison_of_grain_size}

Grain growth in Class 0 protostars and their associated disks has been a subject of significant interest in recent years. Several recent studies, including \citet{Bate2022}, \citet{Kawasaki2022}, \citet{Koga2022},  \citet{Lebreuilly2023} and \citet{Marchand2023} have provided compelling evidence suggesting the possibility of grain growth up to at least few hundred microns in these environments.
\citet{Bate2022} conducted simulations of prestellar core collapse and followed them up to the formation of a rapidly rotating marginally gravitationally unstable first core. 
This represents a stage previous to the formation of a well defined protostellar disk, as the one in IRAS 16293 B, but shows similar physical properties. 
\citet{Bate2022} demonstrated that the combination of enhanced collisional rates and efficient grain growth mechanisms can lead to the formation of substantial dust aggregates in Class 0 disks, growing up to 100\,$\mu$m in this first core stage. 
These results are in line with our possibility to have 100\,\um or even up to 1000\,\um sized particles in a Class 0 disk. 
Similarly, \citet{Kawasaki2022} evolved one-zone non-ideal MHD models of the collapse of dense cores up to densities comparable to those found in our disk model ($n_{\rm g}\gtrsim10^{12}\,$cm$^{-3}$). 
Their calculations focused on the evolution of the grain sizes along the collapse, including the effects of coagulation and fragmentation of silicate dust. 
Their models show that grains can coagulate up to a few 100 microns at the densities of our disk model, and therefore those expected in IRAS 16293 B, and even up to millimetric sizes within the very inner dense regions of the disk. 
These results seem to also be in line with our grain size estimations for the disk in IRAS 16293 B.  
The similar recent work conducted by \citet{Koga2022} simulated the formation of a protostellar disk with mass, radius and age similar to our disk model, i.e., representative of a Class 0 disk. 
In this simulations, they did not include dust coagulation but rather fixed grain sizes over the whole evolution. 
Their results indicate that only the big grains (100-1000\,$\mu$m) are tightly coupled to the gas within the disk, while small grains ($\lesssim$10\,$\mu$m) are partly depleted or swept out form the disk.  
More recently, \citet{Lebreuilly2023} presented hydrodynamical simulation of protostellar collapse and followed the evolution of the grain size distribution up until the formation of a first core, with densities similar to those found by \citet{Bate2022} and in our disk. 
Their results also suggest that grains can coagulate up to a few tens of microns in the the protostellar envelope and up to a few 100\,$\mu$m within the first core.  
The results from \citet{Marchand2023} similarly suggests that grain growth is extremely rapid once a disk is formed. 
Their simulations show how grain sizes can reach more than 100\,$\mu$m in the inner disk within 1000\,yr after disk formation.     
Older predictions for grain sizes in protostellar cores, such as those from \citet{Ormel2009}, \citet{Hirashita2009a} and \citet{Hirashita2013}, suggested that even isothermally collapsing dense cores can achieve coagulation up to 100\,$\mu$m but only if the cloud's dynamical support slows down the collapse beyond its free-fall time and let the grains grow. 
In the case of a collapse faster than the free-fall time ($\lesssim$10$^5$\,yrs), grains should coagulate up to a most a few 10\,$\mu$m. 

Observational constraints from dust emissivity indices do also suggest that grains in protostellar (Class 0) envelopes are bigger than in the ISM and their maximum sizes could be from few 10 to 1000\,$\mu$m \citep{Kwon2009,Agurto-Gangas2019, Miotello2014, Bracco2017, Valdivia2019, LeGouellec2019, Galametz2019, Hull2020, Maureira2022}. 

It is also worth mentioning that a dust population with maximum grain sizes comparable to ISM values ($\lesssim1\,\mu$m; \citealt{MRN}) was not consistent with IRAS 16293 B based on the model-observation comparison presented in \citet{Zamponi2021}. 

The size distribution is yet another source of uncertainty for comparing models to observations. 
Although all the theoretical works mentioned above predict significant grain growth in the envelope around Class 0 protostars and in particular in the high-density material in the disks (e.g., \citealt{Marchand2023}), the resultant distribution of the grain sizes might be different among different studies (e.g., \citealt{Bate2022} and \citealt{Lebreuilly2023}). 
While in \citet{Lebreuilly2023} the resultant distribution resembled a power-law, as the one used here, in \citet{Bate2022} the resultant distribution resembled more of a log-normal. 
This differences can be attributed to how the collision velocities for the grains are calculated (see discussion in \citealt{Lebreuilly2023} and \citealt{Marchand2023}). 
We find that, if we were to consider a log-normal distribution for the grain sizes, the opacities at millimeter wavelengths would slightly increase but still remain within the same order of magnitude as the ones in Fig. \ref{fig:dust_opacities}. 
Such limited variations cannot help in constraining the dust grain size distribution with our current model and thus, our conclusions regarding the maximum grain size remain the same regardless of the adopted size distribution.

\subsection{Grain growth within the soot-line} 
\label{sec:discussion_grain_growth}

The recent VLA observation of FU Ori presented by \citet{Liu2021b} showed that within the inner 10\,au, where T$\gtrsim400$\,K, dust grains have grown to millimeter sizes. 
Since this region is hot enough for evaporation of the water icy mantles, this implies that grain growth was efficient in this "dry" conditions. 
This scenario of grain growth in the inner and hotter regions of protoplanetary disks is also supported by laboratory experiments \citep{Hiroshi2020}.
They show that the stickiness of silicate and carbonaceous grains is enhanced after the removal of quasi-liquid layers of water. 
This process facilitates dust coagulation and supports the formation of planetesimals in dry environments \citep{Pillich2021}. 

The possibility to have millimetric grains can potentially be explained by the high fragmentation velocity $v_{\rm frag}$ found for water-ice-free grains $\gtrsim$10 m s$^{-1}$ \citep{Kimura2015, Gundlach2018, Steinpilz2019, Pillich2021}, larger than previously considered for rocky poorly-sticky grains of 1 m s$^{-1}$ \citep{BlumAndWurm2000}. 
These results on the fragmentation velocity have also been independently found by \citet{Liu2021b} and more recently by \citet{Yamamuro2023}.

\begin{figure*}
    \centering
    \includegraphics[width=0.59375\textwidth]{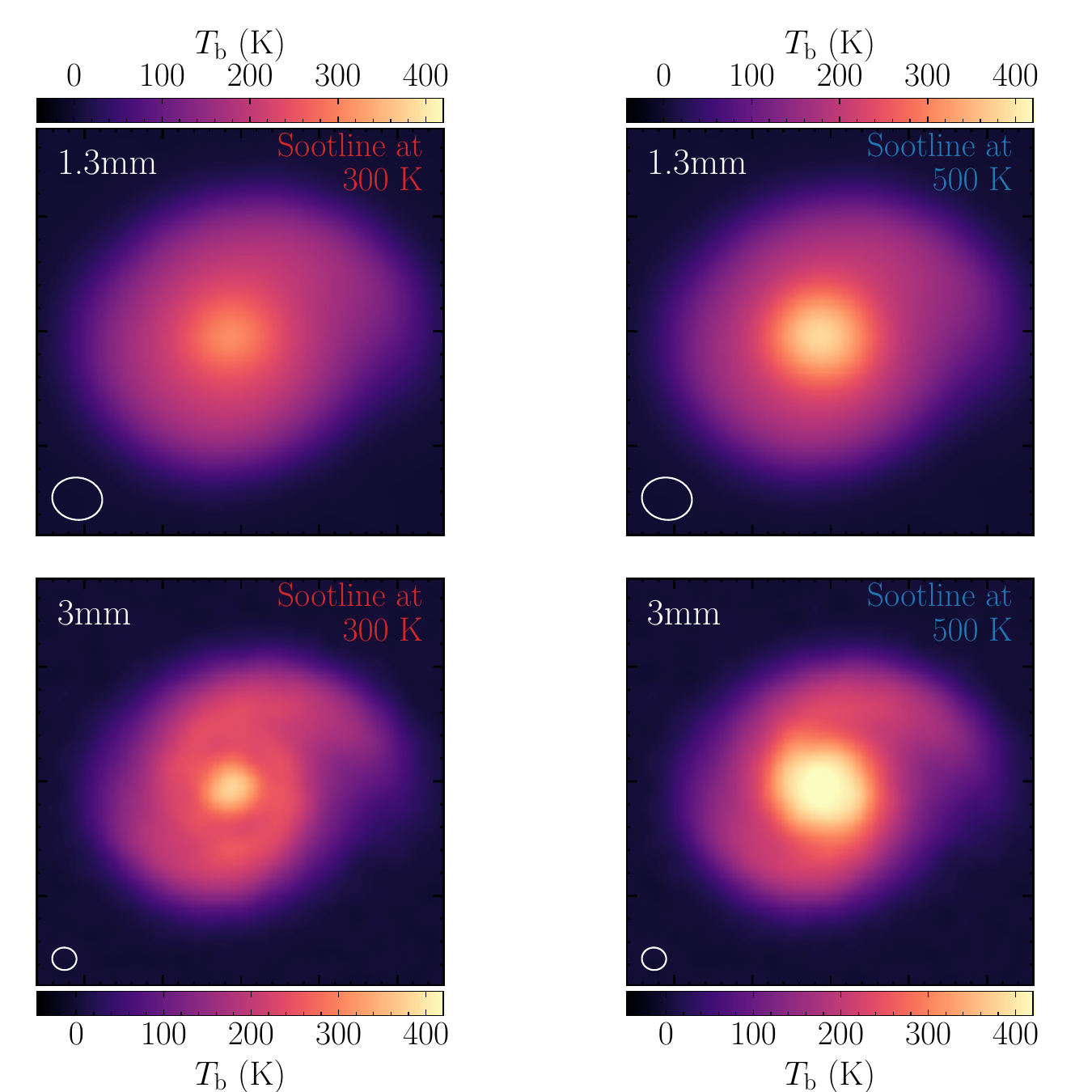}
    \includegraphics[width=0.30625\textwidth]{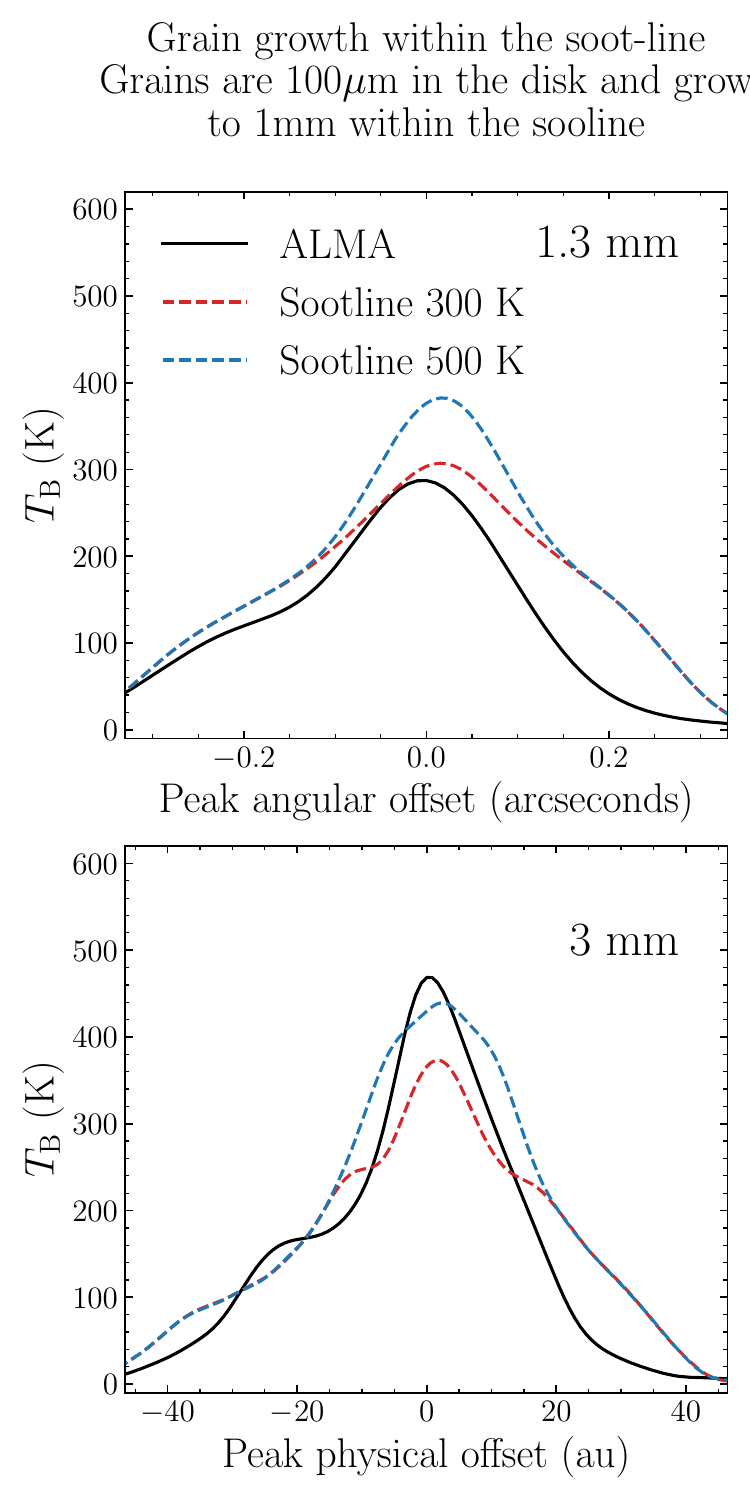}
    \caption{Models with grain growth within the sublimation zone. In these models, the maximum grain size is 100\um outside of the sublimation zone (i.e., in regions where T$\,>300$\,K for the upper panels and T$\,>500$\,K for the lower panels) and 1000\um within it, {with a mixture of silicates, graphites and amorphous carbon (80\% of the carbon)}.}
    \label{fig:grain_growth_soot-line}
\end{figure*}

Motivated by these recent laboratory experimental results, we reproduced synthetic brightness profiles at 1.3 and 3\,mm (similar to Fig. \ref{fig:horizontal_cuts}), accounting for sublimation and grain growth when the temperature is over 300\,K. 
Outside of the sublimation zone the maximum grain size is 100\,$\mu$m and a mixture of silicates, graphites and refractory organics (80\% of the carbon budget).   
Inside of it, we have increased the grain size to 1000\,$\mu$m.  
The results of this test are shown in Fig. \ref{fig:grain_growth_soot-line} for two different sublimation temperatures, 300 and 500 K, at 1.3 and 3 mm. 
We present the synthetic ALMA Stokes I maps, at the resolution of the observations for comparison with IRAS 16293 B. 
The resulting brightness profiles show that the combination of the overestimation in the case of $a_{\rm max}=100\,\mu$m is mitigated by the reduction of flux in the center when $a_{\rm max}=1000\,\mu$m within the sootline. 
We present the results for two different sootlines, to explore the effects of considering the range of sublimation temperatures presented by \citet{Lin2021}. 
The fluxes from models with a 500 K sootline are higher than with 300 K. 
This happens because with a sootline of 500 K the region traced by millimetric grains is more compact than the case with 300 K, this means that the region with higher opacity is smaller in the 500 K case.  Another feature observed in Fig. \ref{fig:grain_growth_soot-line} is the appearance of a gap in the model with sootline at 300 K observed at 3\,mm. 
This is caused by the reduction of the opacity, and is not seen in the model with 500 K because the region with reduced opacity is more compact and less resolved by the observations. 
This shows that observations of such gaps for hot disks could then be the result of changes on the opacity.

Considering grain growth within the sootline results in a better match with the observations than the cases with homogeneous grain sizes. 
The comparison suggests that grain growth up to millimeter sizes could be enhanced within the sootline in this young Class 0 disk, provided the disk densities temperatures and densities in the model are close to the ones present in IRAS 16293B.  

\subsection{If not self-scattering, what is the polarization mechanism in IRAS 16293-2422 B?}
\label{sec:discussion_polarization_mechanisms}
As we have shown in section \ref{sec:self-scat_models}, the emission from IRAS 16293B is too optically thick to produce detectable levels of polarization by self-scattering when considering spherical grains. 
This raises the question of what other mechanisms or grain properties can we consider to explain the observed polarization patterns and fractions. 

Polarization observations are commonly associated with the optically thin emission of magnetically aligned grains \citep{LazarianAndHoang2007a, Andersson2015}. 
However, the emission detected from young embedded disks is likely optically thick \citep{Galvan-Madrid2018, Lin2020, Zamponi2021, Ohashi2023, Lin2023b}.  
In the context of optically thick emission, the polarization observed might also come from magnetically aligned grains, but produced in the form of extinction \citep{Ko2020, Liu2021a}.  
In this case, the differential attenuation of the two orthogonal components of the light E-vectors, results in an excess of polarization along a given axis \citep{Wood1997}. 
This is known as dichroic extinction and occurs when the grains are elongated and the background light is optically thick and almost unpolarized.  
In the context of optically thick embedded disks, this can be produced by foreground (e.g., envelope) elongated grains aligned with their minor axis parallel to the magnetic field lines, which absorbs light preferentially along the grain's major axis and produce a net polarization parallel to the magnetic field lines, as opposed to the optically thin polarized emission. 
Such a polarization mechanism has actually been proposed and detected in a few other Class 0 sources and contradicts the current understanding of magnetic field structures within the optically thick inner regions of the disks \citep{Liu2021a}. 
Similarly, the multiwavelength polarization observations of NGC 1333 IRAS4A, presented by \citet{Ko2020}, show a transition between E-vectors parallel to the magnetic field, traced at 0.87-1.3\,mm, to E-vectors perpendicular to the magnetic field, traced at 6.9-14.1\,mm. 
These results show evidence for a transition between extinction and emission of aligned grains which is determined by an optically thick to thin transition. 
Similarly, polarization by dichroic extinction has been detected within the inner 100\,au of the Class 0 protostar OMC-3/MMS 6 (also known as HOPS-87) in Orion \citep{Liu2021a}, after comparing ALMA and VLA observations. 
Moreover, in the protostar HH212, the polarization observed at a resolution of 14\,au, shows a possible combined contribution of both dichroic extinction and self-scattering \citep{Lee2021}. 
These studies support that dichroic extinction by aligned grains can be an important mechanism to explain the polarization pattern at or close to disks scales in several Class 0 protostars. 
As this mechanism requires that the region near disk scales is optically thick, it would be present preferentially in younger and more massive disks such as those in the Class 0 stage.

In the case of IRAS 16293B, the 1.3\,mm polarization observations from \citet{Sadavoy2018} show azimuthal E-vectors between 30-50\,au and uniform vectors within 30\,au (see Fig. \ref{fig:observations}). 
In \citet{Zamponi2021} and this work we have shown that the emission is optically thick and can reach optical depths well above 100 at 1.3\,mm within the central 10\,au. 
If the polarized emission is produced by dichroic extinction of magnetically aligned grains, this pattern would indicate a toroidal magnetic field. 
Alternatively, the polarization pattern could still be related to the direct emission of magnetically aligned grains close to the disk. 
The magnetic field morphology (traced by B-vectors) could be associated with a poloidal field in that scenario.  
Future modeling and higher-resolution observations can help constraining the contribution of dichroic extinction as well as magnetic field configurations that can best explain the observations in Fig. \ref{fig:observations}.  

Another possible scenario producing azimuthal E-vectors, as those in Fig. \ref{fig:observations}, is the polarization reversal effect, happening when millimetric elongated grains are present in the disk and observed in the Mie regime (i.e, millimetric wavelengths). 
This has been used by \citet{Guillet2020} to explain the azimuthal polarization pattern of [BHB 2007] 11 (the pretzel) in the Pipe nebula, where polarization E-vectors, and so B-vectors, could be aligned with the accretion streamers shown by \citet{Alves2019}. 

Another scenario is that the polarization can be produced by dust alignment, but not necessarily with the magnetic field only.  
Both emission and extinction of light from elongated particles lie on the assumption that dust grains are aligned with a given underlying field.  
This could be either the magnetic, radiation or the velocity field \citep{Gold1952, Wood1997, Lazarian2007}. 
The most commonly accepted mechanisms responsible for the alignment of grains (see \citealt{Andersson2015, Reissl16} or \citealt{Hoang2022} for a review) can be either associated to Radiative Torques (RAT) \citep{LazarianAndHoang2007a} or to supersonic motions, namely, mechanical alignment (MET) \citep{Gold1952, LazarianAndHoang2007b, Kataoka2019, Reissl2023, Hoang2022}.    
In the case of RAT, this can lead to alignment with the radiation field (k-RAT; \citealt{Tazaki2017}) instead of the magnetic field (B-RAT). 
In the case of mechanical alignment this could result in alignment with the magnetic field (B-MET) or the velocity field (v-MET; \citealt{Hoang2022}).  
In the scenarios of alignment with the radiation (k-RAT) or the dust-gas drift velocity field (v-MET or MAT; \citealt{Hoang2018}), the resulting polarization pattern would be azimuthal, as long as the radiation field is assumed to be centrally concentrated and the disk to be Keplerian (e.g., around a protostar), respectively. 
However, the recent work presented by \citet{LeGouellec2023} shows that k-RAT is not the main mechanism responsible for the observed polarization in Class 0 protostars, since it requires high protostellar luminosities radiating on large and rapidly rotating grains, and such grains are commonly settled toward the midplanes of Class 0 disks, shielded from protostellar radiation.    
The case of mechanical alignment depends strongly on the Stokes number, i.e., the degree of spatial coupling between the gas and the dust. 
Whether radiation or mechanical alignment are efficient in Class 0 disks requires further investigation. 

Finally, given our suggestion of large (>100\,$\mu$m) grains present in this Class 0 disk, we find interesting to consider the scenario of grain rotational disruption induced by radiative torques (RATD), introuduced by \citet{Hoang2019} and recently discussed by \citet{LeGouellec2023} and \citet{Reissl2023}.  
According to RATD, rapidly rotating large aligned grains might become rotationally disrupted when they exceed a certain critical angular momentum and would produce a depletion of grown dust.
\citet{LeGouellec2023} showed that RATD might be effective within the outflow cavity regions of Class 0 protostars, for protostellar luminosities above 20\,$L_{\odot}$, and that the densest regions of the disk are well protected against such intense radiative torques. 
These are the region where we suggest dust is grown to above 100 microns in IRAS 16293B. 

\section{Conclusions}
\label{sec:Conclusions}
In this work we have explored the effects the effects of different maximum grain sizes and carbon sublimation in the millimeter continuum emission from a hot and optically thick Class 0 disk, generated from numerical simulations of prestellar core collapse. 
The disk model has successfully reproduced the fluxes of the nearly face on Class 0 disk IRAS 16293 B \citep{Zamponi2021}. 
Hence, we produced synthetic observations of the different cases to compare with multiwavelength (1.3, 3, 9 and 18\,mm) and high-resolution (6 to 44\,au) observations of IRAS 16293 B, including polarization. 
In order to automate the generation of synthetic observations, we have developed a new publicly available tool called Synthesizer, that allows to generate synthetic models from numerical simulations directly from the command line. 
The conclusions of this work are summarized as follows: 
\begin{itemize}
  
    \item For a dust mixture of silicates and graphites, we extended the results of \citet{Zamponi2021}, that used maximum grain sizes $a_{\rm max}$ of 1\,\um and 10\,$\mu$m, and generated opacity tables for maximum grain sizes of up to 100\,\um and 1000\,$\mu$m.
    Peak fluxes increase with wavelength and decrease with $a_{\rm max}$, a feature of hot and optically thick disks. 
    Optical depths range between 130 to 2 from 1.3 to 18 mm, respectively. 
    The high optical depths and positive temperature gradient toward the center also result in extended regions in the disk with $\alpha<2.5$ for all $a_{\rm max}$ and even below 2 for $a_{\rm max}=100\,\mu$m and 1000$\,\mu$m. 
    Predictions from significant grain growth populations, including $a_{\rm max}=1000\,\mu$m are comparable to the observations from IRAS 16293 B at all observed wavelengths. 
    Hence, significant grain growth could be present in this young Class 0 disk. 

    \item Motivated by the high brightness temperatures ($\gtrsim$400 K) observed toward IRAS 16293 B, we explored the scenario of sublimation of solid amorphous refractory carbon at temperatures above 300 K, the so-called sootline. 
    The sublimation results in a local decrease of the optical depth, due to both the reduction of graphite within the grain carbon budget and the evaporated mass reduction. 
    This decrease produces higher fluxes because the emission then traces deeper and hotter layers of the disk. 
    The difference in the fluxes with and without sublimation are small ($<10$\%). 

    \item We also tested the hypothetical case of grain growth within the sootline, motivated by recent laboratory experiments suggesting that dry grains without ice mantes would enhance stickiness and coagulation. 
    We modeled this scenario with an \amax of 100\um at all disk scales outside of the sootline, and with millimetric grains within it. 
    Our results indicate that a combination of both grain sizes can help to provide a better match to the ALMA observations of IRAS 16293 B. 
      
    \item We generated polarization models by self-scattering for three different maximum grain sizes, 10, 100 and 1000\,$\mu$m. The millimetric polarized intensity is highest for the case with $a_{\rm max}=100\,\mu$m at 1.3\,mm, thanks to the high albedo and not extremely high optical depth, as it happens for millimetric grains. 
    However, the predicted polarization fraction produced by self-scattering are very low ($\lesssim$0.5\%), both for edge-on and face-on disks. 
    These low levels of polarization by self-scattering are too low to be consistent with those observed for IRAS 16293 B at both 1.3\,mm and 7\,mm ($2-4$\%). 
       
\end{itemize}
 
Since self-scattering is unlikely to be at the origin of the high-polarization fractions observed toward the Class 0 disk IRAS 16293 B, future modeling is needed to constraint the mechanism responsible for the observed polarization.  
Similar studies toward other embedded disks would be useful to see if this is a general result for  other Class 0 disks. 
Higher-resolution molecular observations will also help to probe if there is enriched carbon chemistry in the inner regions of the disk.

\begin{acknowledgements}
We thank the anonymous referee for the constructive feedback that led to significant improvement of this manuscript. 
We also thank Cornelis Dullemond, Robert Brunngr\"aber, Tommaso Grassi, John D. Ilee, Michael K\"uffmeier and Valentin Le Gouellec for helping building the setup and implementation of this project.   
J.Z., M.J.M., B.Z., D.S. and P.C. acknowledge the financial support of the Max Planck Society. 
H.B.L. is supported by the National Science and Technology Council (NSTC) of Taiwan (Grant Nos. 111-2112-M-110-022-MY3). 
D.S. is supported by an NSF Astronomy and Astrophysics Postdoctoral Fellowship under award AST-2102405.  
This paper makes use of ALMA data from the following projects: 2015.1.01112.S (PI: S. Sadavoy), 2017.1.01247.S (PI: G. Dipierro) and 2016.1.00457.S (PI: Y. Oya). ALMA is a partnership of ESO (representing its member states), NSF (USA) and NINS (Japan), together with NRC (Canada), MOST and ASIAA (Taiwan), and KASI (Republic of Korea), in cooperation with the Republic of Chile.
The Joint ALMA Observatory is operated by ESO, AUI NRAO and NAOJ. 
The National Radio Astronomy Observatory is a facility of the National Science Foundation operated under cooperative agreement by Associated
Universities, Inc. 
This research made use of APLpy, an open-source plotting package for Python \citep{APLpy} hosted at \url{http://aplpy.github.io}.
\end{acknowledgements}

\bibliographystyle{aa}
\bibliography{references}

\newpage

\begin{appendix}

\section{The DustMixer: a command-line tool to generate dust opacities}
\label{app:dustmixer}
The \textsc{DustMixer} reads in tabulated optical constants $n$ and $k$ (also called refractive indices) for a given material. 
It interpolates the entered values over a user-defined wavelength grid and extrapolate if necessary. The extrapolations are done as a constant and a power-law for both the lower and upper wavelength bounds, respectively.  
Then, it uses the BHMIE algorithm \citep{BohrenAndHuffman} to convert these constants into extinction and scattering dust efficiencies $Q_{\rm ext}$ and $Q_{\rm sca}$, and also scattering matrix components, for all wavelengths and a single grain size. 
This algorithm calculates scattering efficiencies by assuming that re-emitted (scattered) photons can be modeled as the  expansion of vector spherical harmonics. This calculation involves solving infinite Riccati-Bessel functions and is highly-computationally expensive. 
The larger the grain size compared to the wavelength, the more terms need to be added in the series and therefore the longer the calculation. 
The \textsc{Synthesizer} employs a multiprocessing scheme to calculate this efficiencies within minutes, for a proper resolution (200 wavelengths and 200 grain sizes, see Fig. \ref{fig:dust_opacities}).  
To finally obtain dust opacities, all single grain size efficiencies are integrated over a power-law size distribution within a minimum $a_{\rm min}$ and maximum $a_{\rm max}$ grain size, for a given power-law slope $q$. 

To combine different dust compositions several mixing methods exists in the literature. 
Some of them directly mix the refractive indices using mixing rules, such as the Brugemann or Maxwell-Garnet rules (see e.g., \citealt{Birnstiel2018DSHARP}), and create a new homogeneous sphere of mixed material to calculate the opacities. 
Other methods use effective medium theories and consider layered dust grains, composed of a core and a mantle (e.g., \citealt{Ossenkopf1991, OssenkopfAndHenning1994}) to calculate the opacities of grains with different core-mantle volume ratios. 
The \textsc{DustMixer} provides support for both mixing techniques. However, in this work we calculated opacities per material and combined them as a sum weighted by their mass fractions (see section \ref{sec:dust_composition}), meaning, different materials coexist within the grid cells and lead to a net opacity, rather than merging into an alloy. 

More details about this tool and its additional functionalities can be found in its public repository.

\section{Polarization models by self-scattering for an edge-on hot disk}
\label{sec:app_polarization_edge-on_disk}
\begin{figure*}
    \centering
    \includegraphics[width=17cm]{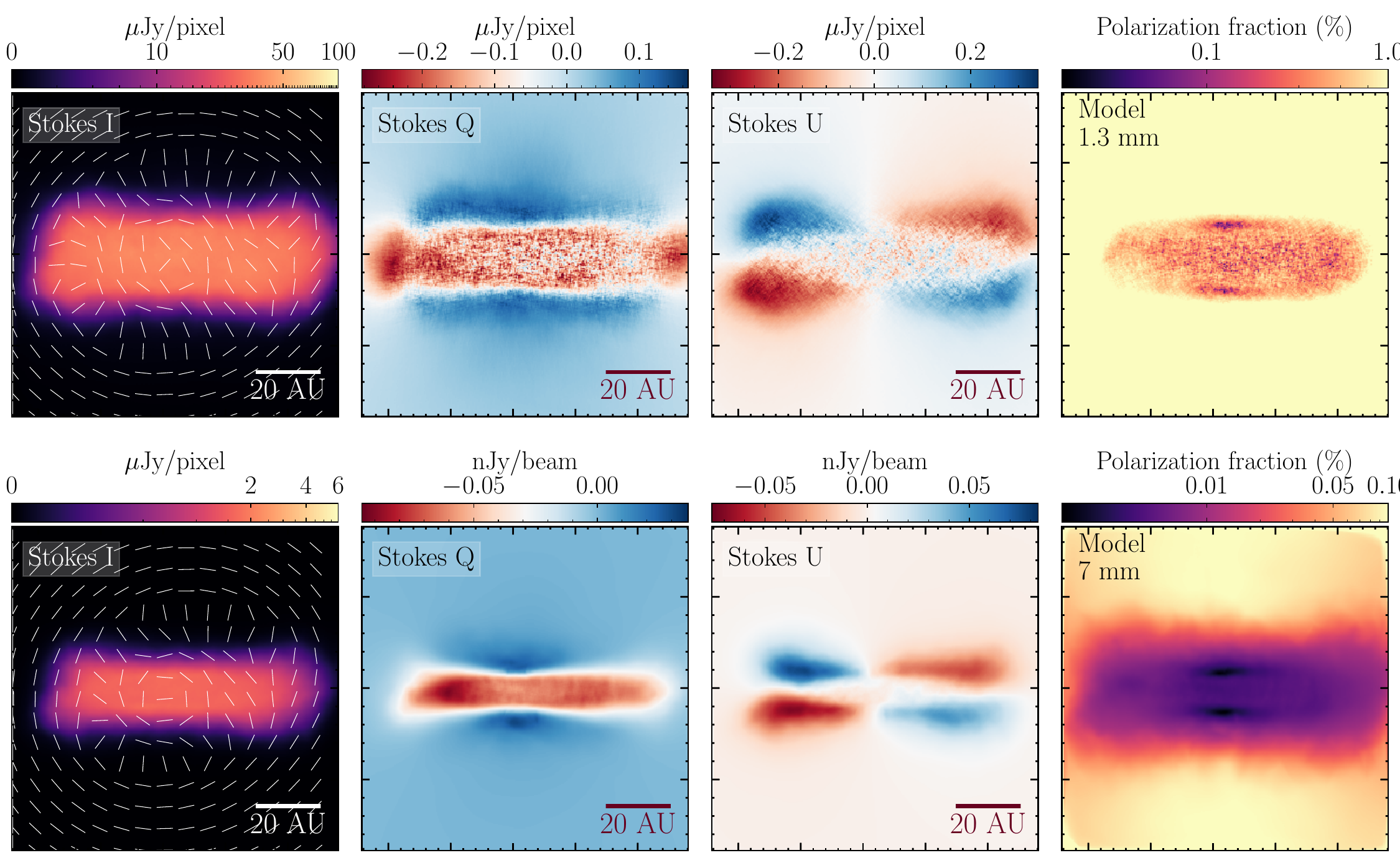}
    \caption{Radiative transfer models of self-scattering at 1.3 (top) and 7\,mm (bottom) for $a_{\rm max} = 100\,\mu$m, shown in the upper and lower row, respectively. This figure is similar to Fig. \ref{fig:self-scat_models_radmc3d} but for an edge-on projection of the same hot and massive disk model. }
    \label{fig:appendix_edge-on_disk}
\end{figure*}

Here we present polarized radiative transfer models produced by self-scattering, similar to those presented in section \ref{sec:polarized_scattering} and shown in Fig. \ref{fig:self-scat_models_radmc3d}, but this time using a edge-on projection for the same disk model. 
Our aim is to provide more global predictions for self-scattering polarization in early embedded Class 0 disks than only the case of the nearly face-on disk on IRAS 16293 B. 
The results of this model are shown in Fig. \ref{fig:appendix_edge-on_disk} for our fiducial dust population of silicates and graphites with $a_{\rm max}=100\,\mu$m, analogous to the case in Fig. \ref{fig:self-scat_models_radmc3d}.  
The polarization fraction of the edge-on models is slightly larger than for the face-on projection ($\lesssim1$\%) but still low to match the ALMA and JVLA observations from Fig. \ref{fig:observations}.

\end{appendix}

\end{document}